\documentclass[twocolumns,a4paper]{aa}
\usepackage{graphicx}
\usepackage{txfonts}
\usepackage{color}
\usepackage{natbib}
\usepackage{hyperref}
\usepackage{auto-pst-pdf}

\usepackage{amsmath}

\def\df{{\rm d}}
\newcommand{\ve}[1]{{\rm\bf {#1}}}

\def\ez{{\bf e_z}}

\def\xas{x^{\ast}\!}
\def\yas{y^{\ast}\!}
\def\zas{z^{\ast}\!}

\def\zda{z^{\dagger}\!}

\begin{document}

\title{Combining magneto-hydrostatic constraints with Stokes profiles inversions.
  I. Role of boundary conditions}
\author{J.M.~Borrero\inst{1} \and A.~Pastor Yabar\inst{1} \and M.~Rempel\inst{2} \and
  B.~Ruiz Cobo\inst{3,4}}
\institute{Leibniz-Institut f\"ur Sonnenphysik, Sch\"oneckstr. 6, D-79104, Freiburg, Germany
\and
High Altitude Observatory, NCAR, P.O. Box 3000, Boulder, CO 80307, USA
\and
Instituto de Astrof{\'\i}sica de Canarias, Avd. V{\'\i}a L\'actea s/n, E-38205, La Laguna, Spain
\and
Departamento de Astrof{\'\i}sica, Universidad de La Laguna, E-38205, La Laguna, Tenerife, Spain
}
\date{Recieved / Accepted}

\abstract{Inversion codes for the polarized radiative transfer equation, when applied to
  spectropolarimetric observations (i.e., Stokes vector) in spectral lines, can be used to infer 
  the temperature $T$, line-of-sight velocity $v_{\rm los}$, and magnetic field $\ve{B}$  as a 
  function of the continuum optical-depth $\tau_{\rm c}$. However, they do not directly provide 
  the gas pressure $P_{\rm g}$ or density $\rho$. In order to obtain these latter parameters,
  inversion codes rely instead on the assumption of hydrostatic equilibrium (HE) in addition 
  to the equation of state (EOS). 
  Unfortunately, the assumption of HE is rather unrealistic across magnetic field lines, causing 
  estimations of $P_{\rm g}$ and $\rho$ to be unreliable.
  This is because the role of the Lorentz force, among other factors, is neglected. Unreliable gas pressure and
  density also translate into an inaccurate conversion from optical depth $\tau_{\rm c}$ to geometrical height $z$.}
{We aim at improving the determination of the gas pressure and density via the application of magneto-hydrostatic  (MHS)
  equilibrium instead of HE.}{We develop a method to solve the momentum equation under MHS
  equilibrium (i.e., taking the Lorentz force into account) in three dimensions. The method is based
  on the iterative solution of a Poisson-like equation. Considering the gas pressure $P_{\rm g}$ and density
  $\rho$ from three-dimensional magneto-hydrodynamic (MHD) simulations of sunspots as a benchmark, we compare
  the results from the application of HE and MHS equilibrium using boundary conditions
  with different degrees of realism. Employing boundary conditions that can be applied to actual
  observations, we find that HE retrieves the gas pressure and density with an error
  smaller than one order of magnitude (compared to the MHD values) in only about 47 \% of the grid points in the
  three-dimensional domain. Moreover, the inferred values are within a factor of two of the MHD values in only about 23 \% of the
  domain. This translates into an error of about $160-200$ km in the determination of the $z-\tau_{\rm c}$ conversion (i.e., Wilson
  depression). On the other hand, the application of MHS equilibrium with similar boundary conditions
  allows determination of $P_{\rm g}$ and $\rho$ with an error smaller than an order of magnitude in 84 \% of the
  domain. The inferred values are within a factor of two in more than 55 \% of the domain. In this latter case, the
  $z-\tau_{\rm c}$ conversion is obtained with an accuracy of $30-70$ km. Inaccuracies are due in equal part to deviations
  from MHS equilibrium and to inaccuracies in the boundary conditions.}{Compared to HE, our new method, based on MHS equilibrium, 
  significantly improves the reliability in the determination of the density, gas pressure, and conversion between geometrical height 
  $z$ and continuum optical depth $\tau_{\rm c}$. This method could be used in conjunction with the inversion of the radiative transfer
  equation for polarized light in order to determine the thermodynamic, kinematic, and magnetic parameters of the solar atmosphere.}

\keywords{Sun: sunspots -- Sun: magnetic fields -- Sun: photosphere -- Magnetohydrodynamics
  (MHD) -- Polarization}
\titlerunning{Magneto-hydrostatic constraints in Stokes profiles inversions}
\authorrunning{Borrero et al.}
\maketitle

\section{Introduction}
\label{sec:intro}

Inversion codes of the radiative transfer equation applied to spectropolarimetric observations of
the solar surface across spectral lines are arguably the most widely used tools to infer the
physical parameters of the solar atmosphere \citep{hector2001review,jc2003review,luis2006review,
  basilio2007review,jc2016review}. These observations correspond to the so-called Stokes vector,
$\varmathbb I(x,y,\lambda)$, where the coordinates $(x,y)$ refer to the solar surface and $\lambda$
is the wavelength. Applied to this kind of observation, inversion codes provide physical parameters such
as the temperature $T$, three-components of the magnetic field $B_{\rm x}, B_{\rm y}$, and $B_{\rm z}$, and so on,
as a function of $(x,y,\tau_{\rm c})$, where $\tau_{\rm c}$ refers to the continuum optical-depth (i.e., far
away from any spectral line). This is possible because scanning in wavelength $\lambda$ is equivalent
to sampling layers located at different optical depths in the solar atmosphere. Most if not all
current inversion codes for the radiative transfer equation, such as SIR \citep{basilio1992sir}, NICOLE
\citep{hector2015nicole}, SPINOR \citep{frutiger2000spinor,vannoort2012decon}, and SNAPI \citep{milic2018snapi},
provide the inferred physical parameters as a function of the continuum optical depth $\tau_{\rm c}$.
This is a consequence of the former being the natural choice to describe the mean-free path of the
photons. It is possible to provide the parameters as a function of the coordinate $z$ by applying
the following relation between  $\tau_{\rm c}$ and the vertical coordinate $z$:

\begin{equation}
  \df \tau_{\rm c} = - \rho \kappa_{\rm c}(T,P_{\rm g})  \df z \;,
  \label{eq:ztau}
\end{equation}

\noindent where $\rho$ is the density and $\kappa_{\rm c}$ is the continuum opacity. The latter is a
nonlinear function of the temperature $T$ and the gas pressure $P_{\rm g}$. To evaluate the equation
above,  $T$, $\rho$, and $P_{\rm g}$ are required. One of these thermodynamic 
parameters (temperature, gas pressure, or density) can be obtained from the other
two by applying a suitable equation of state. There are two main sources of uncertainty when
converting from the $\tau_{\rm c}$-scale to the $z$-coordinate or vice versa. The first is the inaccuracy in the
top boundary condition for the gas pressure $P_{\rm g}(z_{\rm max})$. This has the effect of shifting
the entire $z$-scale upwards or downwards for each atmospheric column (i.e., at fixed $(x,y)$). The second
source of error is the uncertainty in the determination of $T$, $P_{\rm g}$, and $\rho$ elsewhere
outside the upper boundary as a function of $z$, and has the effect of locally stretching or shrinking the $z$ 
spacing between discrete $\tau_{\rm c}$ grid points. 

While the $T$ is retrieved by the inversion code itself, $\rho$ and/or $P_{\rm g}$
must be obtained by other means. This is because these two latter parameters cannot be inferred simultaneously with the temperature 
\citep[see][]{adur2019invz} unless we provide spectral lines of different ionization stages. Unfortunately, this is 
not generally the case and therefore density and gas pressure are instead determined through 
additional constraints, namely the equation of state plus some equilibrium condition. In the case of the aforementioned 
inversion codes this condition is hydrostatic equilibrium. It is clear however that the assumption of hydrostatic 
equilibrium is unreliable in many regions of the solar photosphere, making the retrieved gas pressure and density 
not accurate enough so as to guarantee that the $z$-scale obtained from Equation~\ref{eq:ztau} is trustworthy.\\

Recently, \cite{loptien2018} made use of the null divergence condition of the magnetic field, $\ve{\nabla} \cdot \ve{B}=0$, 
to obtain a $z-\tau_{\rm c}$ conversion where the inferred Wilson depression $z(\tau_{\rm c}=1)$ is within 100 km of the true
Wilson depression. This method however works only for the $\tau_{\rm c}=1$-level and does not attempt to provide 
realistic values for the density and gas pressure.\\

Another possibility would be to circumvent Eq.~\ref{eq:ztau} entirely by working directly in the $z$-scale. 
To that effect, we recently presented a new inversion code \citep[FIRTEZ;][]{adur2019invz} that solves the forward 
and inverse equation for polarized radiative transfer directly in the $z$-scale instead of the 
$\tau_{\rm c}$-scale, thus providing the physical parameters (temperature, magnetic field, etc.) as a function of $(x,y,z)$. 
Unfortunately FIRTEZ also suffers from similar shortcomings to those of the other inversion codes in that the reliability
of the $z$-scale depends on an accurate determination of the density $\rho$ or gas pressure $P_{\rm g}$.
We are then left with only one means of improving the accuracy in the inference of $\rho$ and $P_{\rm g}$
: by dropping the assumption of hydrostatic equilibrium.\\

Early attempts to determine a more accurate $z-\tau_{\rm c}$ conversion based on magnetohydrostatic instead
of hydrostatic equilibrium were by \citet{keller1990},~\citet{solanki1993zw},~\citet{valentin1993zw}, and
\citet{mathew2004zw}. These
works considered cylindrical symmetry however. More recently, \cite{puschmann2010pen} developed a
new method that does not assume any particular symmetry. Unfortunately, the latter method
was not coupled with the inversion code in the sense that the newly retrieved $\rho$ and $P_{\rm g}$ were
not fed back into the inversion algorithm to fit the observed Stokes vector $\varmathbb I(x,y,\lambda)$.
\cite{puschmann2010pen} noticed for instance that the new gas pressure would appreciably change the continuum
in the intensity profiles. Building up from their idea we aim at developing a new method to obtain more
realistic densities and gas pressures based also on magnetohydrostatic equilibrium but in such a fashion
that the resulting values can be fed back into the inversion code we have developed \citep[FIRTEZ;][]{adur2019invz}.\\

In the present work we assume that the inversion of Stokes profiles provides the temperature $T(x,y,z)$ and magnetic
field $\ve{B}(x,y,z)$ as given by three-dimensional magnetohydrodynamics (MHD) simulations of sunspots
(Section~\ref{sec:3dsimul}). Furthermore, here we focus only on the effects of the boundary conditions.
Errors in the determination of the temperature and magnetic fields via the inversion of Stokes profiles,
along with the effects of the limited spatial resolution in the observations, will be addresses in future work.
Under this premise, we study the reliability of the inference of the density $\rho$
and gas pressure $P_{\rm g}$ using hydrostatic equilibrium (Section~\ref{sec:hydeq}) and magnetohydrostatic
equilibrium (Section~\ref{sec:mhseq}) and compare our results with the more realistic density and gas
pressure from the MHD simulations.  The accuracy of the presented methods in the determination
of the $z-\tau_{\rm c}$ conversion (Eq.~\ref{eq:ztau}) is assessed in Section~\ref{sec:discussion}. The limitations
of the method and possible ideas as to how to combine the method presented here with inversion codes for the radiative
transfer equation are addressed in Section~\ref{sec:limitations}. Finally, Section~\ref{sec:conclu}
summarizes our findings.\\

\section{3D nongray MHD simulations}
\label{sec:3dsimul}

\begin{figure}[!h]
\begin{center}
\includegraphics[width=9cm]{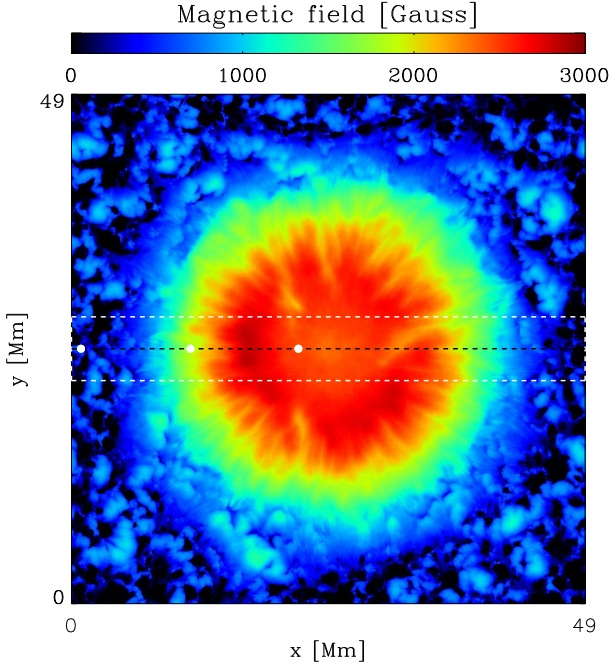}
\caption{Magnetic field $B$ from the sunspot simulation at a height of $448$~km from the upper boundary.
  The rectangular box in white-dashed lines is the region employed for our study.\label{fig:bfimhd}}
\end{center}
\end{figure}

\begin{figure}[!h]
\begin{center}
\includegraphics[width=9cm]{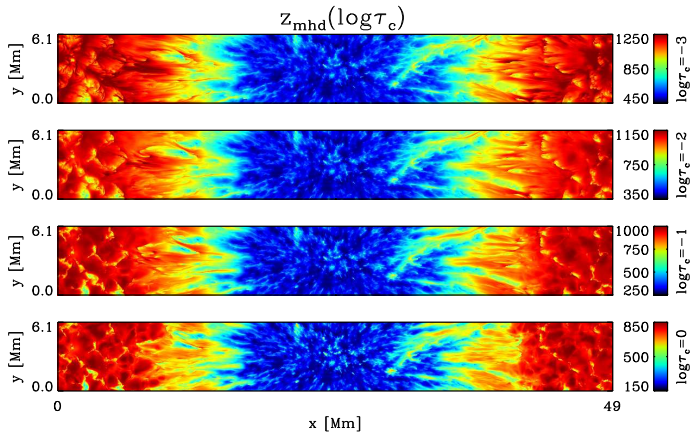}
\caption{Geometrical height $z$ for four different optical-depth levels. From bottom to top:
  $\log\tau_{\rm c}=[0,-1,-2,-3]$.\label{fig:zmhd}}
\end{center}
\end{figure}

Our investigations are based on a nongray three-dimensional MHD simulation of a sunspot
following the setup described in \citet{rempel2012mhd}. The resulting sunspot models
cover $49.152\times 49.152\times 6.144\,\mbox{Mm}^3$, and were computed using gray 
radiative transfer and different grid resolutions. To obtain them, 
we restarted a nongray simulation from the model with
$16\times 16\times 12\,\mbox{km}^3$ resolution in \citet{rempel2012mhd} and evolved it for an additional
15 minutes with nongray radiative transfer at a higher resolution of $12\times 12\times 8$~km. At this
resolution the domain has a size of $4096\times 4096\times 768$ grid points. Along the third dimension
(i.e.,  direction of gravity) only the upper 192 grid points are needed. These are enough to cover the entire
photosphere, which is defined as the region above the $\tau_{\rm c}=1$-level (i.e., continuum optical depth
unity level), both in the granulation and umbra, including the Wilson depression. In the granulation
surrounding the sunspot the $\tau_{\rm c}=1$-level is located around $z \approx 1000$ km. For illustration
purposes a map of the magnetic field $B$ at a fixed height of $448$~km from the top boundary (i.e.,
close to $\tau_{\rm c}=1$ in the surrounding granulation) is shown in Figure~\ref{fig:bfimhd}. The rectangular
region limited by the white-dashed lines is the one employed in our study. It encompasses
$4096 \times 512 \times 192$ grid points covering $49.152 \times 6.144 \times 1.536\,\mbox{Mm}^3$.
Along the $x$-axis it runs over the umbra, penumbra, and surrounding granulation. Figure~\ref{fig:zmhd}
shows $z(x,y,\tau_{\rm c})$ for four optical-depth levels (from bottom to top): $\log\tau_{\rm c}=[0,-1,-2,-3]$.
This figure can be viewed as the Wilson depression at different $\tau_{\rm c}$-levels. For instance, the
$z(\log\tau_{\rm c}=0)$-level is located approximately 500-600 km deeper in the umbra ($x\approx 25$ Mm) than in
the surrounding granulation ($x \approx 0$ Mm). These four optical-depth levels have been chosen
because they represent what is commonly considered as the photosphere.\\

\section{Hydrostatic equilibrium}
\label{sec:hydeq}

Hydrostatic equilibrium implies that the gas pressure is stratified only due to gravity according to:

\begin{equation}
{\bf \nabla} P_{\rm g,hyd} = -\rho_{\rm hyd} \ve{g} ,
\label{eq:hydeqvec}
\end{equation}

\noindent where $\rho$ is the density and $\ve{g}=g\ez$ is the acceleration due to gravity. On the solar
surface $g =2.7414 \times 10^{4}$ cm~s$^{-2}$ (cgs units are employed throughout this paper). The
vertical $z$-component of the equation above translates into\\

\begin{equation}
\frac{\partial P_{\rm g,hyd}}{\partial z} = - \rho_{\rm hyd} g .
\label{eq:hydeqz}
\end{equation}

This is a first-order ordinary differential equation that needs only one boundary condition (BC). This BC is typically
set at the uppermost boundary $z_{\rm max}$, so that Eq.~\ref{eq:hydeqz} is integrated backwards. Because
the gas pressure tends to decay exponentially\footnote{This case represents the exact solution
for the isothermal case with constant ionization.} with $z$, the alternative procedure, that is, setting the BC
at $z_{\rm min}$ and integrating upwards, is usually avoided so as to prevent negative values of $P_{\rm g,hyd}$.
To study the accuracy to which hydrostatic equilibrium can determine the gas pressure and density we have
solved Eq.~\ref{eq:hydeqz} along the $z$-direction employing a fourth-order Runge-Kutta method for each grid
point $(x,y)$ in the horizontal plane in the three-dimensional domain of the MHD simulation (Sect.~\ref{sec:3dsimul}), and using in
each case two different boundary conditions for $P_{\rm g,hyd}(z_{\rm max})$. The first BC takes the gas pressure
on the uppermost horizontal plane to be exactly identical to that from the MHD simulations in
Sect.~\ref{sec:3dsimul}: $P_{\rm g,hyd}(x,y,z_{\rm max}) = P_{\rm g,mhd}(x,y,z_{\rm max})$. The second BC takes the
gas pressure on the uppermost horizontal plane to be axisymmetric and equals to:

\begin{eqnarray}
  \begin{split}
    \log \tilde{P}_{\rm g}(\xi,z_{\rm max}) =  & -0.401-2.679 \xi +6.353 \xi^2 \\ & -2.788 \xi^3 + 0.303 \xi^4
    \label{eq:bctophydro2}
  \end{split}
,\end{eqnarray}

\noindent where $\xi=r/R_{spot}$ is the normalized radial distance from the  center of the Sunspot. To understand
where this equation comes from we refer the reader to Section~\ref{subsec:realmhs}. At this point we simply state that Eq.~\ref{eq:bctophydro2} results in a value for the gas pressure at $z_{\rm max}$ of about $0.4$ and
$160$ dyn cm$^{-2}$ in the umbral center ($\xi=0$) and surrounding granulation ($\xi=2$), respectively.
For comparison purposes we note that the three-dimensional MHD simulations yield typical values for the gas
pressure at $z_{\rm max}$ of the order of $10^2-10^3$ dyn cm$^{-2}$ and $10^{-1}-1$ dyn cm$^{-2}$ in the granulation
and umbra, respectively.\\

\begin{figure*}
\begin{center}
\includegraphics[width=16cm]{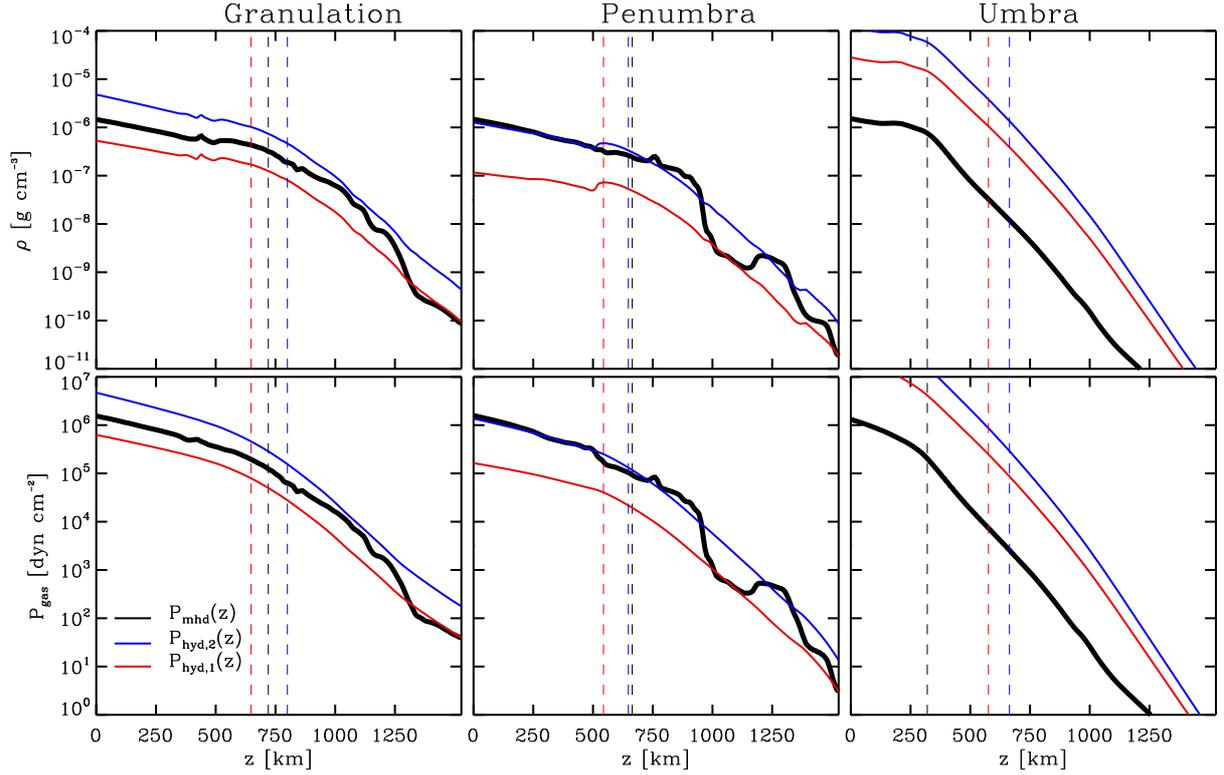}
\caption{{\it Top panels}: Density as a function of the geometrical height $z$ (i.e., vertical coordinate)
  for three spatial $(x,y)$ locations corresponding to the sunspot umbra (right), penumbra (middle),
  and surrounding granulation (left). These locations are indicated by white circles in Fig.~\ref{fig:bfimhd}.
  {\it Bottom panels}: Same as top panels but for the gas pressure. Solid black curves correspond to the actual
  values from the three-dimensional MHD simulation (Sect.~\ref{sec:3dsimul}), whereas colored lines
  are the hydrostatic results using the two boundary conditions described in the text:
  $P_{\rm hyd,1}$ (red) and $P_{\rm hyd,2}$ (blue)\label{fig:hydeq}. The vertical dashed lines
  indicate the location of the $z(\tau_{\rm c}=1)$-level (i.e., Wilson depression).}
\end{center}
\end{figure*}

For the sake of simplicity, results obtained with the aforementioned boundary conditions will be
referred to as $P_{\rm hyd,1}$ and $P_{\rm g,hyd,2}$, respectively. After obtaining the solution for the hydrostatic
gas pressure $P_{\rm g,hyd}$, the hydrostatic densities $\rho_{\rm hyd,1}$ and $\rho_{\rm hyd,2}$ are obtained by
applying the equation of state for ideal gases using the temperature from the MHD simulations:\\

\begin{equation}
\rho = \frac{u}{k} \frac{\mu}{T_{\rm mhd}} P_{\rm g} \,
\label{eq:eos} 
,\end{equation}

\noindent where $T_{\rm mhd}$ is the temperature taken from the MHD simulations, $u=1.66053902 \times 10^{-24}$ g is the
atomic mass unit, and $k=1.38064852 \times 10^{-16}$ erg~K$^{-1}$ is Boltzmann's constant. The mean molecular
weight $\mu$ is itself a function of the temperature and is determined solving the Saha and Boltzmann equations
\citep[][Ch.~3]{mihalas1970} self-consistently for $92$ atomic species. Figure~\ref{fig:hydeq} shows the density
(top panels) and gas pressure (bottom panels) as a function of the vertical $z$-axis for three grid points located in the sunspot
umbra (right), penumbra (middle), and granulation surrounding the sunspot (left). These
grid points are indicated in Fig.~\ref{fig:bfimhd} as white-filled circles. In solid black
we depict the actual values from the MHD simulations (Sect.~\ref{sec:3dsimul}), whereas
in solid red and solid blue we show the results after applying hydrostatic equilibrium (Eq.~\ref{eq:hydeqz})
using the two aforementioned boundary conditions: $P_{\rm g,hyd}(x,y,z_{\rm max})=P_{\rm g,mhd}(x,y,z_{\rm max})$
($P_{\rm hyd,1}$; red), and $P_{\rm hyd,2}(x,y,z_{\rm max})$ given by Eq.~\ref{eq:bctophydro2} (blue; $P_{\rm hyd,2}$).
By construction, solid black and red curves for the gas pressure (bottom) and density
(top) meet at $z_{\rm max}$. The vertical dashed lines indicate the location of the $z(\tau_{\rm c}=1)$-level
(i.e., Wilson depression).\\

As can be seen, the inferred pressure $P_{\rm g,hyd}$ and density $\rho_{\rm hyd}$ stratification as a
function of $z$, as well as the $z(\tau_{\rm c}=1)$-level, are highly dependent on the upper boundary
condition $P(z_{\rm max})$. Moreover, the match between the hydrostatic values and the ones from the MHD
simulations is in general poor, with discrepancies as large as one order of magnitude in the granulation
and penumbra, and up to two orders of magnitude in the umbra (Fig.~\ref{fig:hydeq}; right panels).
Of particular interest is also the fact that, whereas
in the MHD simulations (black-solid curves) the gas pressure can increase or decrease with increasing $z$
(bottom-middle panel around $z \approx 1100$ km), in the hydrostatic case only
$\partial P_{\rm g,hyd}/\partial z < 0$ is allowed so as to avoid negative densities.\\

It can be argued that it should be possible to adapt the upper boundary condition (i.e., consider it as a
free parameter) so as to improve the match between the hydrostatic solutions (solid red/blue curves) and the
magnetohydrodynamic (solid black) values. This is equivalent to a change in the integration constant
in Eq.~\ref{eq:ztau} mentioned in Section~\ref{sec:intro}. While this is certainly a possibility, this idea cannot
be applied to real observations because in this case we do not have any information about the
real pressure and density, and therefore there is nothing to match $P_{\rm g,hyd}$ and $\rho_{\rm hyd}$ to.\\

It is important to point out that neither of the two boundary conditions employed above are fully compatible
with hydrostatic equilibrium. The reason is that hydrostatic equilibrium is not only represented by the
$z$-derivative of the gas pressure (Eq.~\ref{eq:hydeqz}), but also by the $x$ and $y$-derivatives in
Eq.~\ref{eq:hydeqvec}: $\partial P_{\rm g,hyd} / \partial x = \partial P_{\rm g,hyd} / \partial y=0$. This
implies that $P_{\rm g,hyd}$ is constant in planes of fixed $z$. This condition is not verified by either
the solid blue or the red curves in Fig.~\ref{fig:hydeq}, as at a given $z$ the gas pressure varies
horizontally (i.e., it is different in the granulation, penumbra, umbra, etc). Indeed, it is clear that 
as long as the temperature is also a function of $(x,y)$, hydrostatic equilibrium cannot be maintained 
in three dimensions. This occurs because if $\partial P_{\rm g,hyd} / \partial x = \partial P_{\rm g,hyd} / \partial y=0$ 
then the same applies (through Eq.~\ref{eq:hydeqz}) to the density: 
$\partial \rho_{\rm hyd} / \partial x = \partial \rho_{\rm hyd} / \partial y=0$. Therefore,
the application of the equation of state (Eq.~\ref{eq:eos}) directly yields: 
$\partial T / \partial x = \partial T / \partial y=0$.\\

\begin{figure*}
\begin{center}
\begin{tabular}{cc}
\includegraphics[width=8cm]{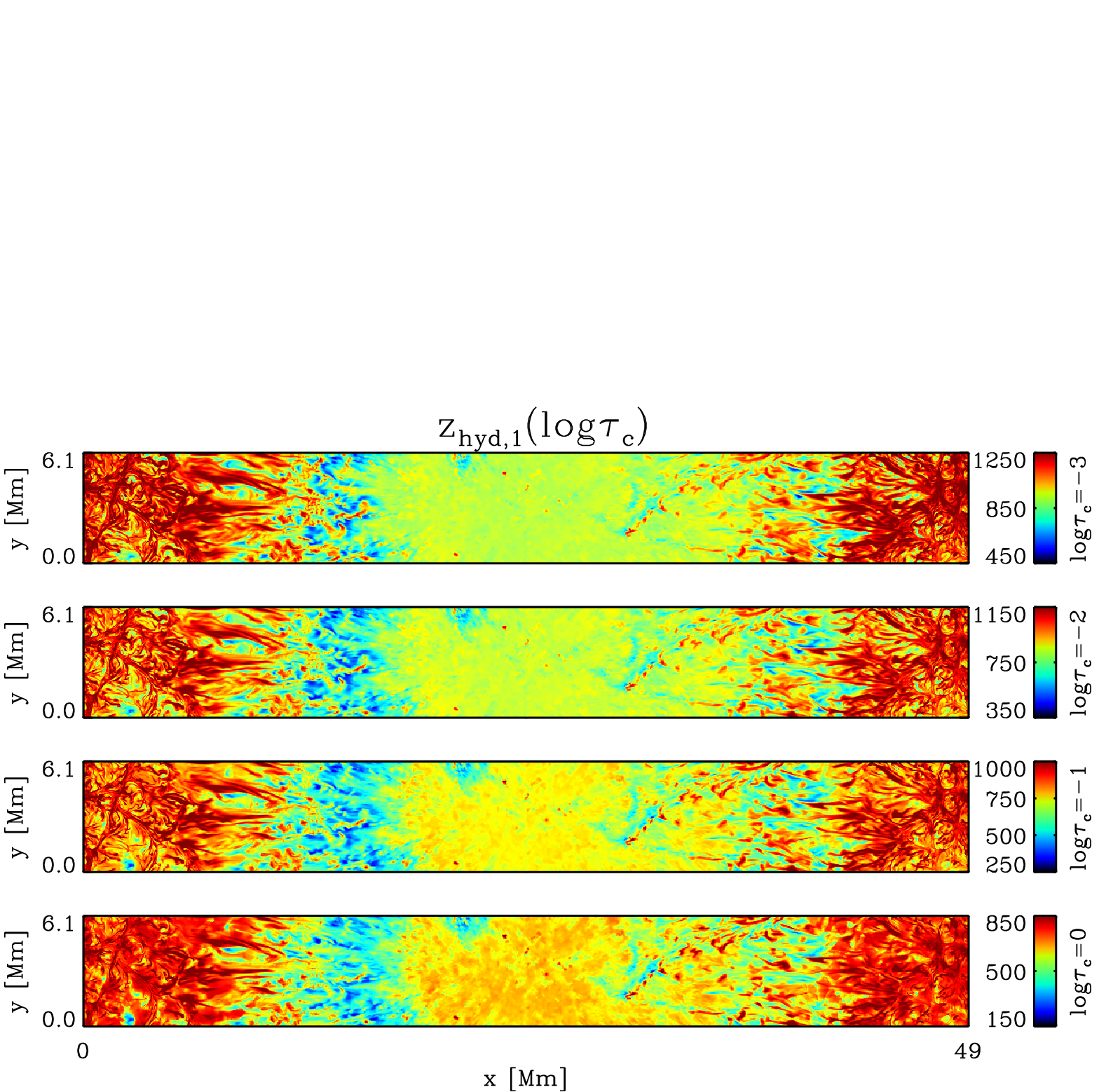} &
\includegraphics[width=8cm]{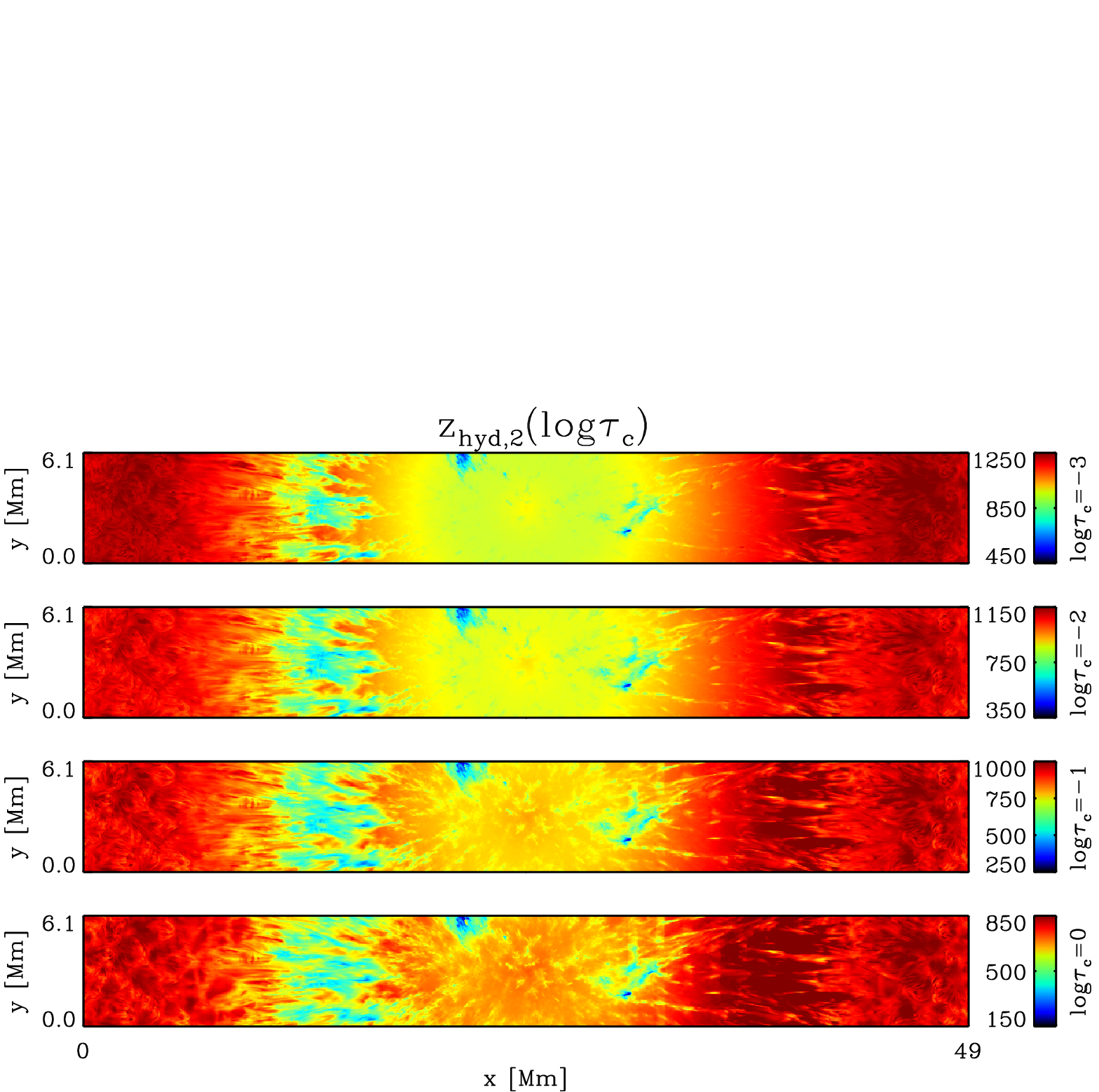}
\end{tabular}
\caption{{\it Left panels}: Two dimensional maps of $z(x,y,\tau_{\rm c})$ at four different $\log\tau_{\rm c}$-levels (from
  bottom to top): $0,-1,-2,-3$ using hydrostatic equilibrium and the boundary condition
  $P_{\rm g,hyd}(x,y,z_{\rm max}) = P_{\rm g,mhd}(x,y,z_{\rm max})$. {\it Right panels}: Same as on the left but
  using the boundary condition where $P_{\rm g,hyd}(x,y,z_{\rm max})$ is given by Eq.~\ref{eq:bctophydro2}.\label{fig:zhydeq}}
\end{center}
\end{figure*}

Thus far, we have demonstrated that the gas pressure and density inferred through hydrostatic equilibrium 
signficantly differ from the values in the MHD simulations (by as much as two orders of magnitude). The
question now is how these inaccuracies translate into $z-\tau_{\rm c}$ conversion as given by Eq.~\ref{eq:ztau}.
This is addressed in Figure~\ref{fig:zhydeq}, where we display $z(x,y,\tau_{\rm c})$ at $\log\tau_{\rm c}=0,-1,-2,-3$ (from bottom
to top). Left and right panels in this figure correspond to the results obtained with the first and second boundary
condition, $P_{\rm hyd,1}$ and $P_{\rm hyd,2}$, respectively. As can be seen by comparison with Fig.~\ref{fig:zmhd},
the agreement between the hydrostatic solutions and the MHD simulations is rather poor. In particular the Wilson
depression between the umbra and surrounding granulation is only about 150 km, whereas in the MHD simulations is closer to
500-600 km. In addition, there is a very strong asymmetry between the penumbra on either side of the umbra,
in particular when employing the $P_{\rm hyd,2}$
boundary condition given by Eq.~\ref{eq:bctophydro2}. While this asymmetry does not appear in the maps
of the Wilson depression of the MHD simulations (Fig.~\ref{fig:zmhd}) it does indeed originate in the simulations,
with the left-side penumbra having a stronger magnetic field than the right side due to the presence
of a very elongated penumbral filament that protrudes into the umbra in a way that resembles a light bridge.
This renders the thermal structure of the left and right sides slightly different. Under hydrostatic equilibrium
these different temperatures immediately translate into different pressure and density stratifications with $z$
and thus also a different $z-\tau_{\rm c}$ conversion.\\

\section{Magneto-hydrostatic equilibrium}
\label{sec:mhseq}

Under magneto-hydrostatic (MHS) equilibrium, the momentum equation takes the following form:\\

\begin{equation}
{\bf \nabla} P_{\rm g,mhs} = -\rho_{\rm mhs} \ve{g} +\frac{1}{4\pi} ({\bf \nabla} \times \ve{B}) \times \ve{B} ,
\label{eq:mhseqvec}
\end{equation}

\noindent which is obtained by adding the Lorentz force term to
Eq.~\ref{eq:hydeqvec}. This is a system of three first-order partial differential equations.
To avoid dealing with such a system of equations we take the divergence of Eq.~\ref{eq:mhseqvec}
and transform it into a single second-order partial differential equation:

\begin{equation}
  \nabla^2 P_{\rm g,mhs} = -g \frac{\partial \rho_{\rm mhs}}{\partial z} +\frac{1}{4\pi} {\bf \nabla} \cdot
        [({\bf \nabla} \times \ve{B}) \times \ve{B}] .
\label{eq:mhspois}
\end{equation}

This equation is now a Poisson-like equation that can be solved provided that the right-hand-side is
known. As mentioned in Section~\ref{sec:intro} we are assuming throughout this paper that the inversion of
Stokes profiles provides the temperature $T$ and magnetic field, $B_{\rm x}, B_{\rm y}, B_{\rm z}$, as a function
of $(x,y,z)$. However we are still lacking the knowledge of the density $\rho_{\rm mhs}(x,y,z)$, which is
in fact one of our unknowns. To circumvent this problem we propose an initial density distribution
$\rho^0_{\rm mhs}$, which is then used to determine the right-hand side of Eq.~\ref{eq:mhspois}. We then
solve this equation employing the \emph{fishpack}\footnote{Fishpack library can be downloaded here
  \url{https://www2.cisl.ucar.edu/resources/legacy/fishpack}} library \citep{fishpack1975}. This yields
a gas pressure $P^0_{\rm g}(x,y,z)$ which, along with the already known temperature $T(x,y,z)$,
is used to obtain a new density $\rho^1_{\rm mhs}$ via the equation of state (Eq.~\ref{eq:eos}). Here,
$\rho^1_{\rm mhs}$ should improve compared to our original estimation $\rho^0_{\rm mhs}$. We then iterate the entire procedure
until convergence is achieved, which we define as being the point at which neither the gas pressure nor the density change significantly
in several consecutive iterations. In all our tests this occurs within 20-30 iterations.\\

Equation~\ref{eq:mhspois} is a second-order partial differential equation, thus requiring two boundary
conditions per dimension. In our case we employ Dirichlet boundary conditions at all six planes
surrounding the domain: $P_{\rm g,mhs}(\xas,y,z)$, $P_{\rm g,mhs}(x,\yas,z)$, $P_{\rm g,mhs}(x,y,\zas)$, where $\xas$
refers to both $x_{\rm min}$ and $x_{\rm max}$, and likewise for $\yas$ and $\zas$.

\subsection{Best-case scenario}

We now consider a \emph{best-case scenario} in which we assume that all values of the gas pressure at
the boundaries in the three-dimensional domain indicated by the white box in Fig.~\ref{fig:bfimhd} are identical
to the MHD values:

  \begin{equation}
    \left\{
    \begin{tabular}{c}
      $P_{\rm g,mhs}(\xas,y,z) = P_{\rm g,mhd}(\xas,y,z)$ \\ $P_{\rm g,mhs}(x,\yas,z) = P_{\rm g,mhd}(x,\yas,z)$ \\
      $P_{\rm g,mhs}(x,y,\zas) = P_{\rm g,mhd}(x,y,\zas).$
    \end{tabular}\right.
    \label{eq:bcmhd}
  \end{equation}

  Further we assume that on the right-hand side of Eq.~\ref{eq:mhspois} the density
  is also given by the values from the MHD simulations. If in the employed simulations
  (see Sect.~\ref{sec:3dsimul}) the additional terms in the momentum equation that are ignored
  by the magneto-hydrostatic equilibrium, such as the time-derivative of the velocity, and the advection
  and viscous terms
  \citep[see Sect~2.2][\footnote{\url{https://ediss.uni-goettingen.de/handle/11858/00-1735-0000-0006-B556-9}}]{voegler2003},
  are negligible, then the gas pressure and density resulting from solving Eq.~\ref{eq:mhspois} should be very similar to
  the values in the MHD simulations. In other words, this test can be considered
  as a study of how close the MHD simulations are to magneto-hydrostatic equilibrium. Gas pressure and density
  obtained with this test are henceforth referred to as $P_{\rm mhs,1}$ and $\rho_{\rm mhs,1}$, respectively.\\

\subsection{Practical scenario}
\label{subsec:realmhs}

\begin{figure}
    \begin{center}
      \includegraphics[width=8cm]{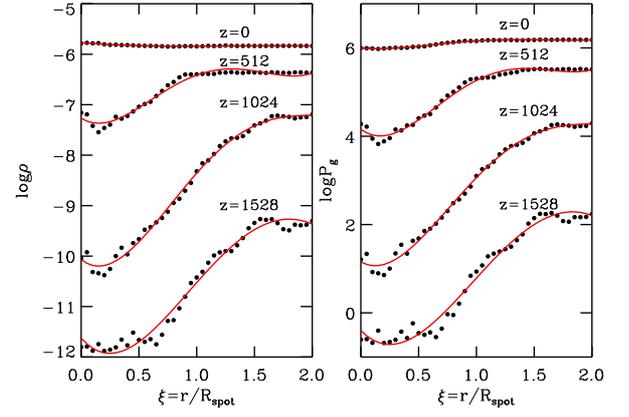}
      \caption{Logarithm of the density (left) and gas pressure (right) as a function of the normalized 
        radial distance to the center of the  sunspot $\xi=r/R_{\rm spot}$ at four different vertical heights 
        $z=0,512,1024,1528$ km. Larger $z$ values correspond to higher atmospheric layers. Black circles correspond
        to the azimuthal- or $\phi$-averages of the MHD simulations (Sect.~\ref{sec:3dsimul}). Red curves are obtained
        through fourth-order polynomial approximations to the black circles. \label{fig:extatm}}
    \end{center}
  \end{figure}

  We subsequently performed \emph{a more practical test}, where we do not assume that the upper/lower and
  side boundary conditions, or the knowledge of the initial density for the iteration of
  Eq.~\ref{eq:mhspois}, are given by the MHD simulations (Eq.~\ref{eq:bcmhd}). Instead we
  employ empirical boundary conditions that can be applied to actual observations. These
  boundary conditions result from polynomial approximations interpolated over the angular
  average of the three-dimensional magnetohydrodynamic simulations at different radial distances
  from the center of the  sunspot. To this end we first convert from Cartesian $(x,y,z)$ to cylindrical $(\xi,\phi,z)$
  coordinates in our three-dimensional domain. Here $\xi$ is the normalized radial distance from the
  center of the  sunspot: $\xi = r / R_{\rm spot}$. We then perform $\phi$-averages (annular) of the logarithm of the gas
  pressure and density from the MHD simulations for $(\xi,z)$ pairs. Finally, we fit fourth-order
  polynomials as a function of $\xi$ at each geometrical height $z_j$. Mathematically,

  \begin{eqnarray}
    \log \tilde{P}_{\rm g}(\xi,z_j) \simeq \sum\limits_{k=0}^{k=4} a_{kj} \xi^k\\ 
    \log \tilde{\rho}(\xi,z_j) \simeq \sum\limits_{k=0}^{k=4} b_{kj} \xi^k
  ,\end{eqnarray}
  
  \noindent where $\tilde{P}_{\rm g}$ and $\tilde{\rho}$ refer to the annular averages. 
  Examples of such polynomial fits are shown in Figure~\ref{fig:extatm} for four different
  heights $z=0,512,1024,1528$ km. As it can seen, Eq.~\ref{eq:bctophydro2} can be obtained
  from the equation above by simply taking $z_j=z_{\rm max}=1528$ km. These polynomials allow us
  to recreate the gas pressure and density at any position $(r,z)$
  in the three-dimensional domain. Hereafter we refer to these values as \emph{interpolated} or \emph{int}
  for short. We employ these polynomial approximations
  to build the initial density $\rho^0=\rho_{\rm int}(r,z)$ used to iterate the solution of Eq.~\ref{eq:mhspois},
  as well as the gas pressure at the boundaries:

  \begin{equation}
    \left\{
    \begin{tabular}{c}
      $P_{\rm g,mhs}(\xas,y,z) = P_{\rm g,int}(\xas,y,z)$ \\ $P_{\rm g,mhs}(x,\yas,z) = P_{\rm g,int}(x,\yas,z)$ \\
      $P_{\rm g,mhs}(x,y,\zas) = P_{\rm g,int}(x,y,\zas).$
    \end{tabular}\right.
    \label{eq:bcext}
  \end{equation}

  The described procedure implies that the boundary conditions for the gas pressure and initial density
  are axisymmetric. We emphasize that the interpolated values (solid red lines in Fig.~\ref{fig:extatm})
  thus obtained are usually within 20\% of the mean values (black circles). However, when compared with the actual
  values from the three-dimensional MHD simulations at individual grid points, differences of an order of magnitude or more are common.
  It is in this sense that we declare this test to be suitable for real observations since it does not require
  accurate knowledge of the boundary conditions. We now have all the ingredients needed to solve Eq.~\ref{eq:mhspois}
  iteratively and obtain the gas pressure and density that are consistent with magneto-hydrostatic equilibrium and are 
  consistent with the temperatures inferred from the inversion of Stokes profiles: $T_{\rm inv}(x,y,z)$. Results
  obtained with these boundary conditions are referred to as $P_{\rm mhs,2}$ and $\rho_{\rm mhs,2}$.\\
  
\begin{figure*}
    \begin{center}
      \includegraphics[width=16cm]{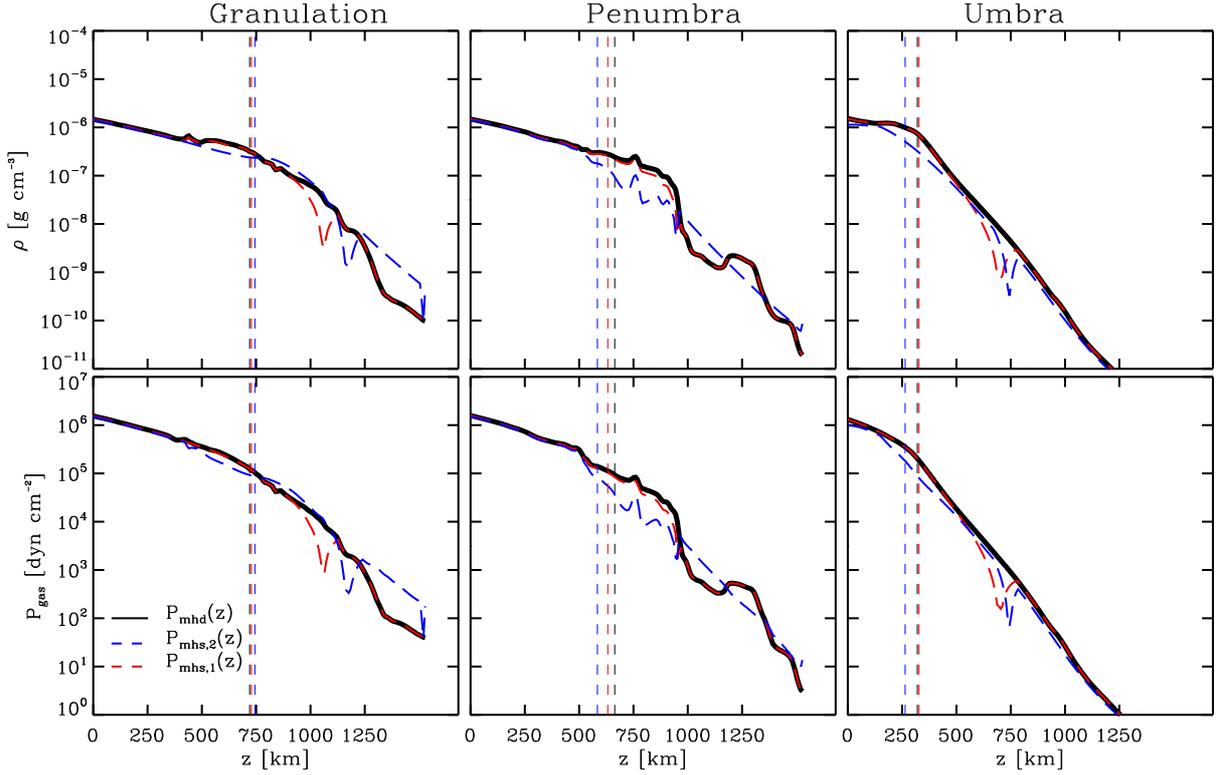}
      \caption{Same as Fig.~\ref{fig:hydeq} but showing results from magneto-hydrostatic equilibrium
        using different boundary conditions: \emph{ideal case scenario} (dashed red lines) where
        boundary conditions are identical to the MHD simulations, and \emph{practical scenario} 
        (dashed blue lines) where boundary conditions are given by the interpolated model (Eq.~\ref{eq:bcext}). Original values
        from MHD simulations (Sect.~\ref{sec:3dsimul}) are displayed in solid black lines.\label{fig:mhseq}}
    \end{center}
  \end{figure*}

  Examples of the retrieved density and gas pressure in the two scenarios we have just described, for the same three
  spatial locations as in Fig.~\ref{fig:hydeq} (see also white circles in Fig.~\ref{fig:bfimhd}), are displayed in
  Figure~\ref{fig:mhseq}. They are indicated by the dashed red and dashed blue lines
  for $P_{\rm mhs,1}$ and $P_{\rm mhs,2}$, respectively. It can be readily seen that the agreement between
  these new results and the ones from the MHD simulations (solid black lines) is not only very good, but is also far better than the
  hydrostatic case (solid red and solid blue lines in Fig.~\ref{fig:hydeq}). Of particular interest is the fact
  that now, the gas pressure $P_{\rm g,mhs}$ can increase with increasing $z$ without involving negative densities.
  It is also worth noting that the derived gas pressure and density depend much less
  on the boundary conditions in the magneto-hydrostatic case than in the hydrostatic one. This occurs because in
  the magneto-hydrostatic case the pressure stratification depends strongly on the Lorentz
  force (second term on the right-hand side of Eq.~\ref{eq:mhspois}). Of particular importance are the $x$ and $y$ variations
  of the magnetic pressure and tension, which couple the results in the horizontal direction, thereby enabling
  the derived values to quickly forget the boundary condition a few grid points away from the upper boundary in the 
  $z$-direction.\\

   \begin{figure*}[ht]
     \begin{center}
       \begin{tabular}{cc}
         \includegraphics[width=8cm]{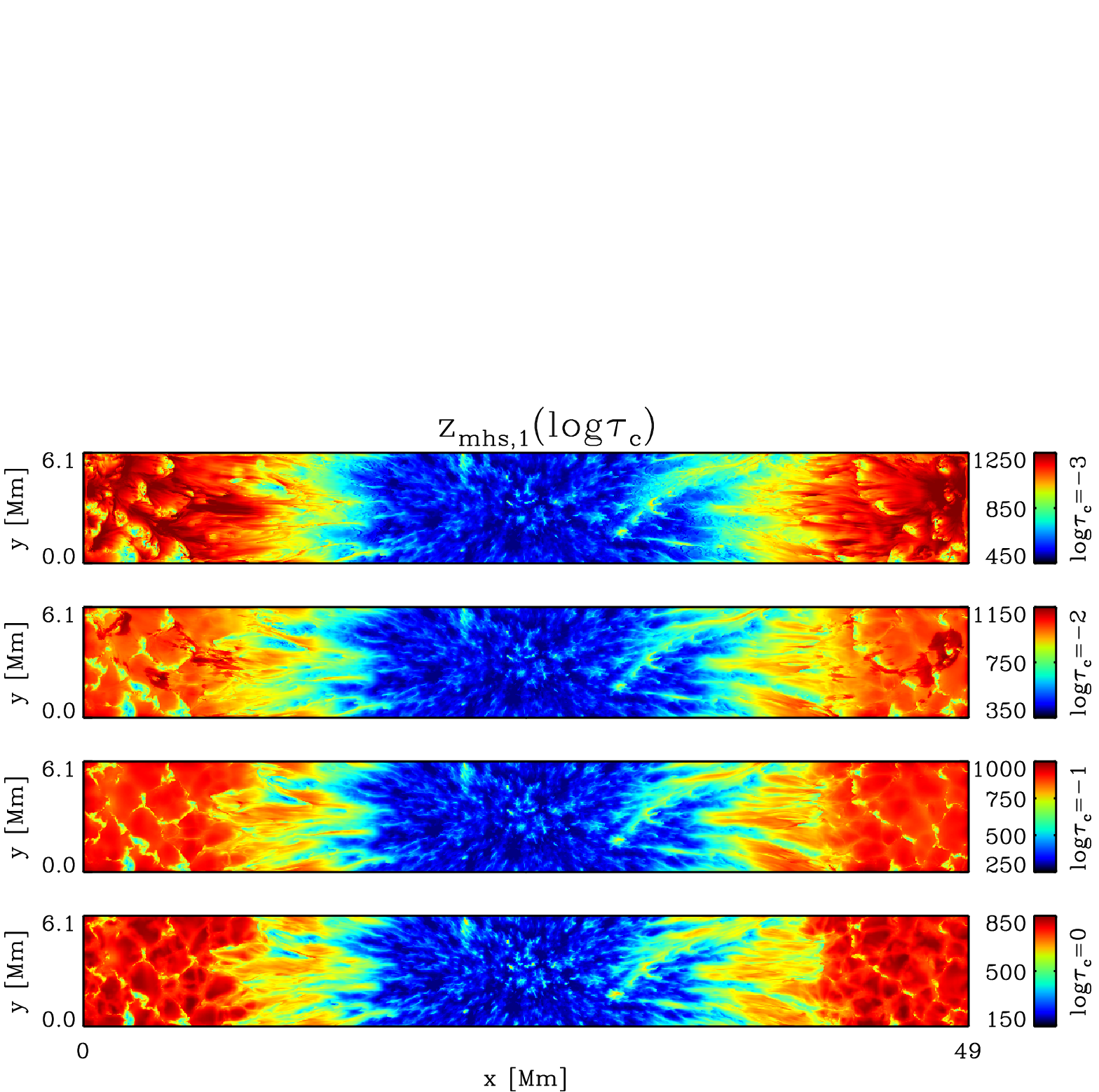} &
         \includegraphics[width=8cm]{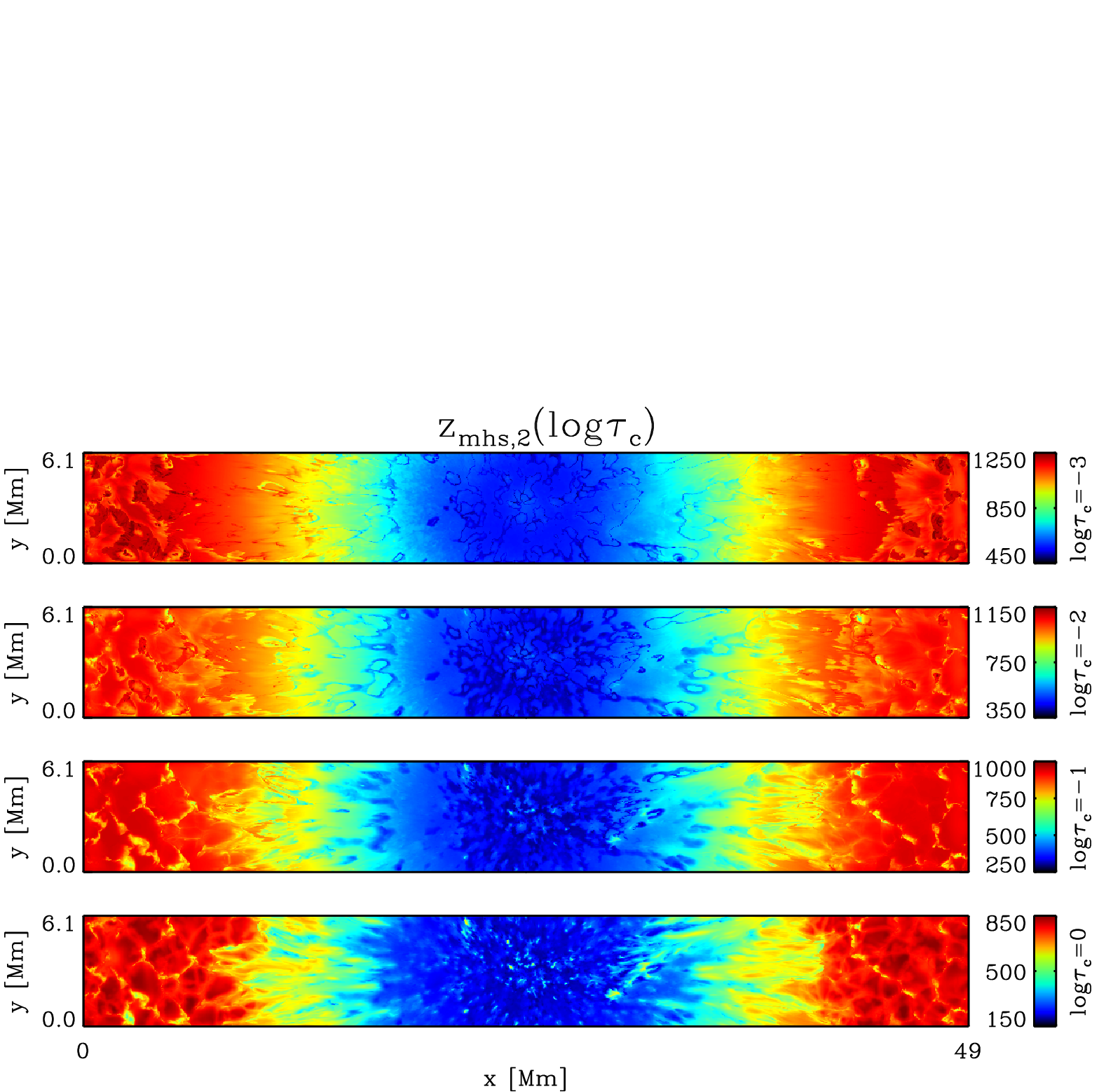} \\
       \end{tabular}
       \caption{Same as Fig.~\ref{fig:zhydeq} but employing magneto-hydrostatic equilibrium
         and different boundary conditions: \emph{ideal case scenario} where boundary conditions
         are identical to the MHD simulations (left panels), and \emph{practical scenario} where boundary
         conditions are given by the interpolated model (right panels; Eq.~\ref{eq:bcext}).\label{fig:zmhseq}}
     \end{center}
   \end{figure*}
   
  Using the gas pressure $P_{\rm g,mhs}$ and density $\rho_{\rm mhs}$ from the previous two tests, and assuming that
  the inversion of Stokes profiles correctly retrieves the temperature from the simulations $T_{\rm mhd}$, we now employ
  Eq.~\ref{eq:ztau} to determine the optical-depth scale $\tau_{\rm c}$. Figure~\ref{fig:zmhseq} shows
  the geometrical height $z$ for the location of the optical-depth values of $\log\tau_{\rm c}=0,-1,-2,-3$ (i.e., Wilson
  depression at different optical depth levels). Again, results are remarkably good and much better than
  the hydrostatic case shown in Fig.~\ref{fig:zhydeq}. Of particular interest is the disappearance of the
  asymmetry between the left and right penumbral sides that was present in the hydrostatic case. Under magneto-hydrostatic
  equilibrium the gas pressure does not depend only on the thermal stratification but also on the Lorentz force. This
  compensates the different temperatures at either side of the umbra, yielding a Wilson depression
  in better agreement with the MHD simulations (Fig.~\ref{fig:zmhd}). It is worth pointing out that
  $z(\log\tau_{\rm c}=-3)$ with interpolated boundary conditions for the gas pressure (Eq.~\ref{eq:bcext}; see 
  uppermost-right panel in Fig.~\ref{fig:zmhseq}) features a very axisymmetric distribution. This occurs because 
  $\log\tau_{\rm c}=-3$ is close to the upper boundary, and therefore results at this layer can be conditioned by the
  axisymmetic boundary conditions imposed there.\\

\section{Discussion}
\label{sec:discussion}

In the above sections we describe two methods to obtain the gas pressure and density
that are consistent with hydrostatic (Sect.~\ref{sec:hydeq}) and magneto-hydrostatic (Sect.~\ref{sec:mhseq})
equilibrium using different boundary conditions. We have seen that the inferred
$P_{\rm g}$ and $\rho$ are qualitatively much closer to the MHD values in the magneto-hydrostatic case
than in the hydrostatic one. The same applies to the $z-\tau_{\rm c}$ conversion. In this section
we present a more quantitative study of the reliability in the inference of the gas pressure
and density as well as of the reliability in the $z-\tau_{\rm c}$ conversion.\\

\begin{figure*}
  \begin{center}
    \begin{tabular}{cc}
      \includegraphics[width=8cm]{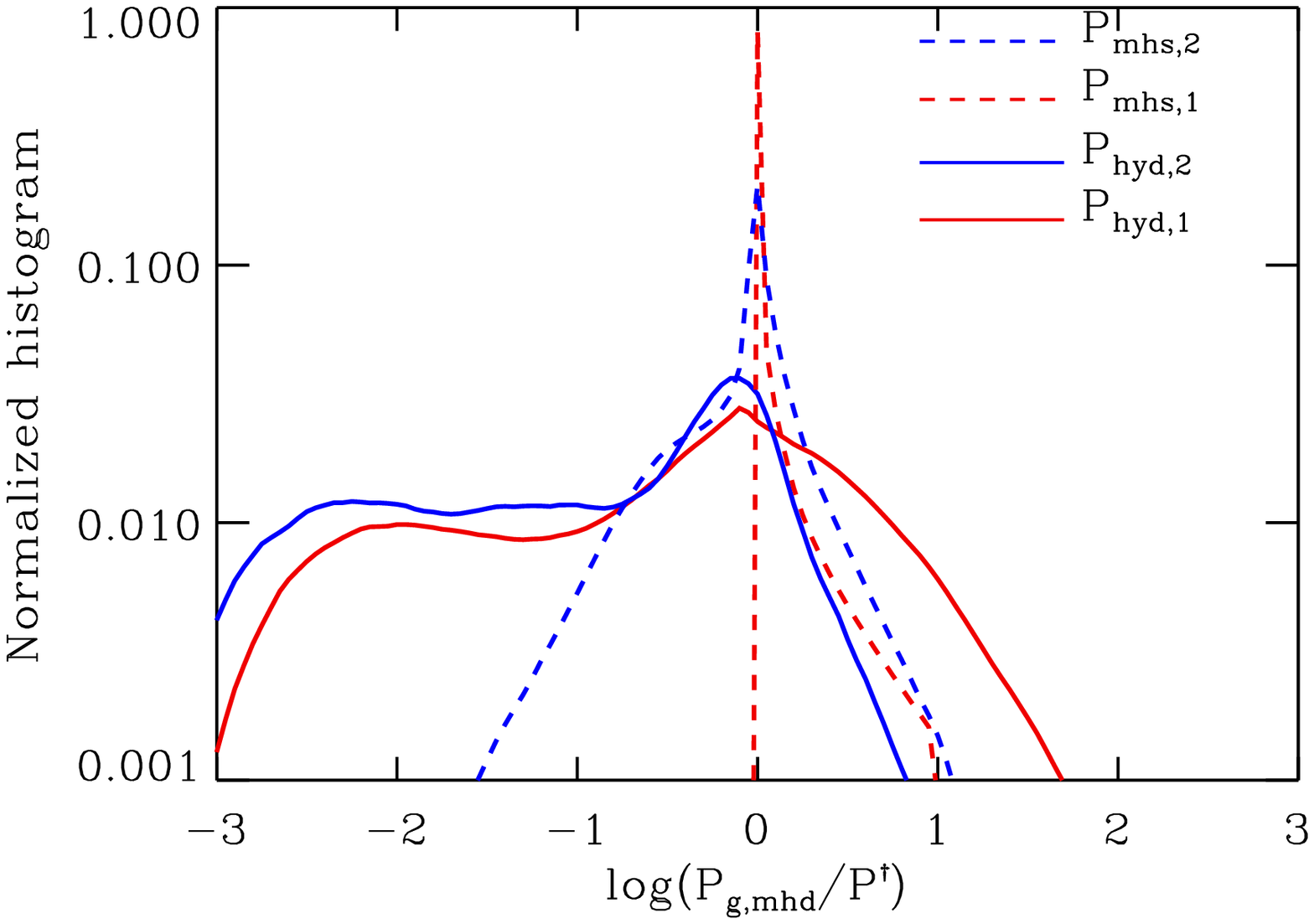} &
      \includegraphics[width=8cm]{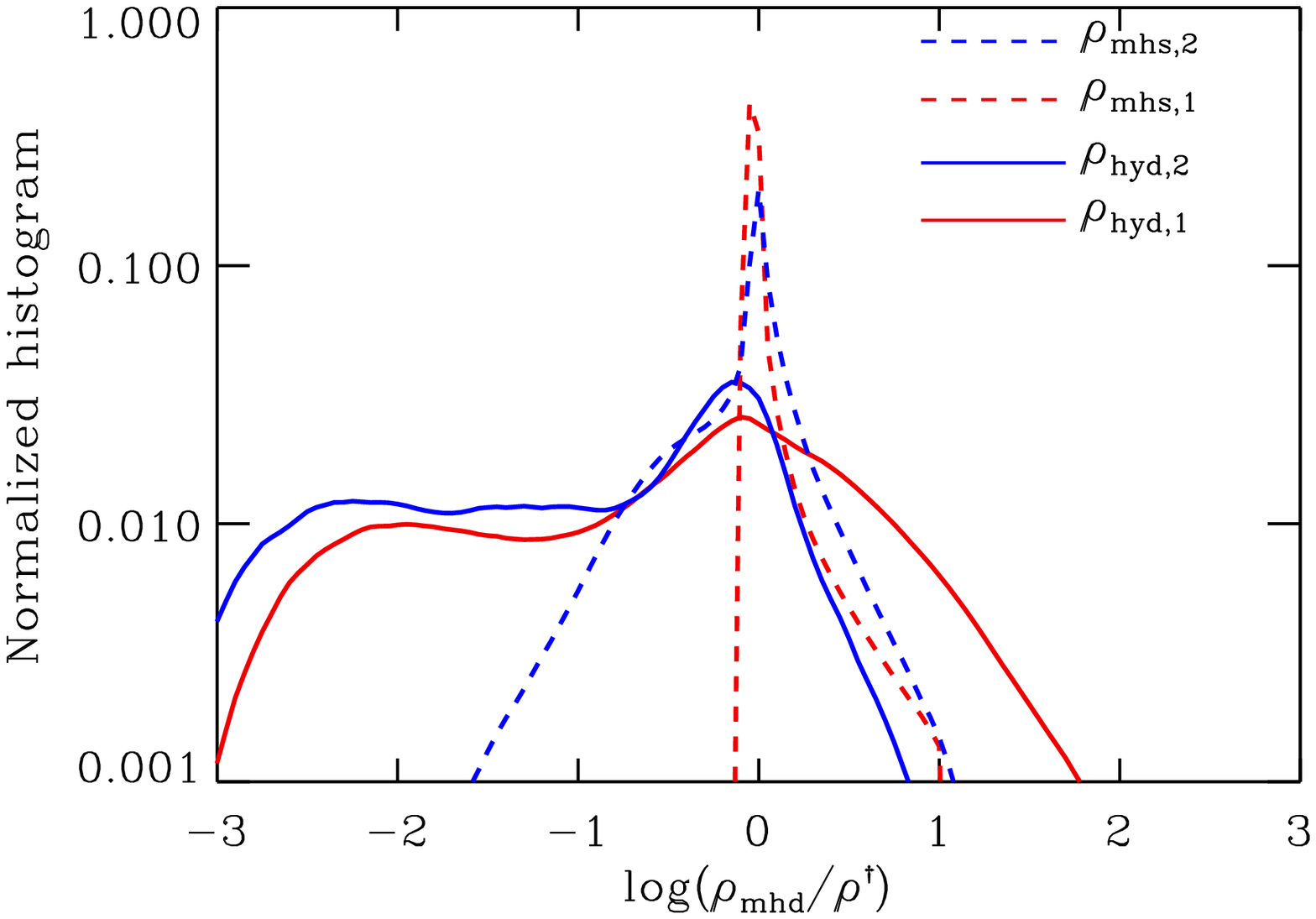}
    \end{tabular}
    \caption{Histograms of the logarithm of the quotient $P_{\rm mhd}/P^{\dagger}$ (left) and $\rho_{\rm mhd}/\rho^{\dagger}$
        (right) in the entire three-dimensional domain, where ${\dagger}$ indicates the values obtained in the four
        different tests we have carried out. Solid colored curves represent the inferences in the
        hydrostatic case (red for boundary condition $P_{\rm hyd,1}$; blue for boundary condition $P_{\rm hyd,2}$),
        and dashed coloured curves display the magneto-hydrostatic case (red for boundary condition $P_{\rm mhs,1}$;
        blue for boundary condition $P_{\rm mhs,2}$). See text for details.\label{fig:predenhist}}
    \end{center}
\end{figure*}

Figure~\ref{fig:predenhist} displays the histograms of the following quantities: $\log(P_{\rm g,mhd}/P^{\dagger})$
(left-panels) and $\log(\rho_{\rm mhd}/\rho^{\dagger})$ (right-panels). Here, $P^{\dagger}$ and $\rho^{\dagger}$ stand for
the gas pressure and density obtained in the different tests carried out in this paper: hydrostatic equilibrium
using boundary condition $P_{\rm hyd,1}$ (solid red line), same but with $P_{\rm hyd,2}$ as boundary condition
(Eq.~\ref{eq:bctophydro2}; solid blue line), magneto-hydrostatic equilibrium with $P_{\rm mhs,1}$ boundary conditions
(Eq.~\ref{eq:bcmhd}; dashed red line), and magneto-hydrostatic equilibrium with $P_{\rm mhs,2}$ boundary conditions (Eq.~\ref{eq:bcext}; dashed blue
line). To estimate the reliability in the determination of the gas pressure and density we determine,
from these histograms, the percentage of points inside the three-dimensional domain where the inferred gas pressure
and density are within one order of magnitude from those in the MHD simulations: $\| \log(P_{\rm g,mhd}/P^{\dagger})\| \le 1$
and $\| \log(\rho_{\rm mhd}/\rho^{\dagger})\| \le 1$. This is equivalent to obtaining the area of each histogram
between the abscissa values $[-1,1]$. Likewise we determine the percentage of points where the inferred gas pressure
and density are within a factor of two from those in the MHD simulations: $\| \log(P_{\rm g,mhd}/P^{\dagger})\| \le 0.3$
and $\| \log(\rho_{\rm mhd}/\rho^{\dagger})\| \le 0.3$. Results for the four tests carried out
in this paper are summarized in Table~\ref{tab:tabpreden}. We discuss here only those results where
we employed boundary conditions that can be applied to real observations: $P_{\rm hyd,2}$ (blue solid-line
in Fig.~\ref{fig:predenhist}) and $P_{\rm mhs,2}$ (blue-dashed line in Fig.~\ref{fig:predenhist}).
Under the assumption of hydrostatic equilibrium the inferred gas pressure and density are within one order of magnitude,
and within a factor of two of the correct values, in only about 47 and 23\%, respectively, of the grid points in
the three-dimensional domain. In the case of magnetohydrostatic equilibrium these numbers increase to about 84 and
55\%, respectively. This latter represents a huge improvement over the hydrostatic case and it certainly opens
the possibility for inversion codes of the polarized radiative transfer equation to infer for the first time, via
magneto-hydrostatic constraints, reliable values of the gas pressure and density in the solar atmosphere.\\

\begin{table*}
\begin{center}
\caption{Summary of results in the determination of gas pressure and density\label{tab:tabpreden}}
\begin{tabular}{ccccc|cc}
          &        &         & $\| \log(P_{\rm g,mhd}/P^{\dagger})\|$ & & $\| \log(\rho_{\rm mhd}/\rho^{\dagger})\|$ & \\
Equation  & $z$-BC & $xy$-BC & \% $\le 1$ & \% $\le 0.3$ & \% $\le 1$ & \% $\le 0.3$ \\
  \hline
  Hydrostatic & mhd & na & 43.4 & 17.5 & 42.6 & 17.0 \\
  Hydrostatic & interpol & na & 46.6 & 23.2 & 46.2 & 22.7 \\
  MHS & mhd & mhd & 98.7 & 93.8 & 99.5 & 91.8 \\
  MHS & interpol & interpol & 83.9 & 55.2 & 83.5 & 54.8 \\
  \hline
\end{tabular}
\end{center}
\tablefoot{$z$-BC: boundary conditions on the uppermost and lowermost XY planes. $xy$-BC: boundary
  conditions on the sides (XZ and YZ planes) of the three-dimensional domain; \emph{mhd} indicates that
  the BC is taken directly from MHD simulations; \emph{interpol} means the BC is taken as an interpolation
  over averaged MHD simulations; na: not applicable.}
\end{table*}

\begin{figure}
\begin{center}
\includegraphics[width=8cm]{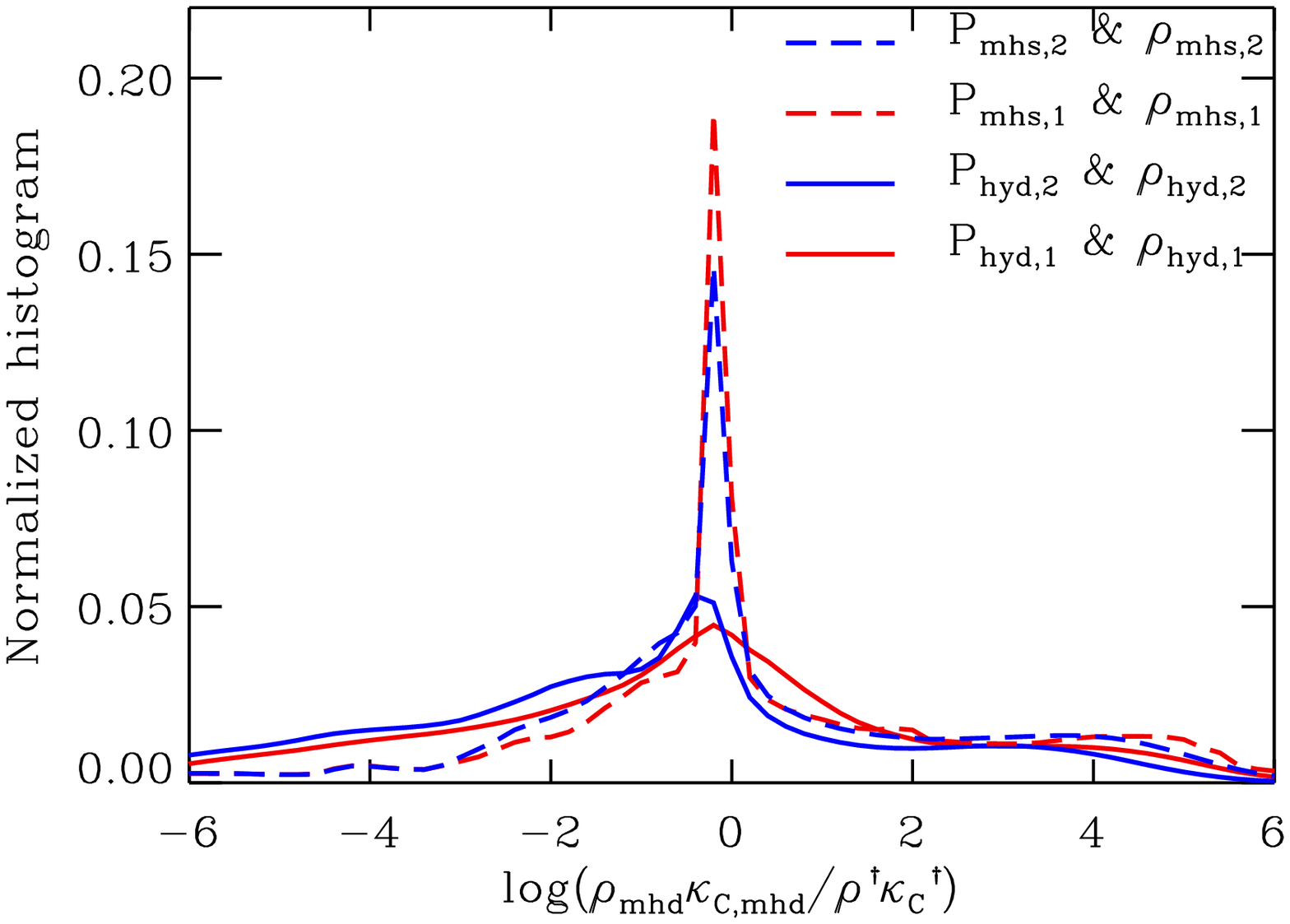}
\caption{Same as Fig.~\ref{fig:predenhist} but for the logarithm of the quotient 
$\rho_{\rm mhd} \kappa_{\rm mhd}/\rho^{\dagger} \kappa_{\rm c}^{\dagger}$
(see Eq.~\ref{eq:dtaudz}) in the entire three-dimensional domain.\label{fig:dtaudz}}
\end{center}
\end{figure}

We now turn our attention to the $z-\tau_{\rm c}$ conversion. As mentioned in Section~\ref{sec:intro}
one of the main sources of uncertainty here is the accuracy in the determination of $T$, $P_{\rm g}$, 
and $\rho$ elsewhere outside the upper boundary as a function of $z$, which has the effect of locally 
stretching or shrinking the $z$ spacing between discrete $\tau_{\rm c}$ grid points. This effect is
determined by the $\df\tau_{\rm c}/\df z$ derivative given by Eq.~\ref{eq:ztau}. To study it
we plot the following ratio in Figure~\ref{fig:dtaudz}:

\begin{eqnarray}
\log \left(\frac{[\df\tau_{\rm c}/\df z]_{\rm mhd}}{[\df\tau_{\rm c}/\df z]^{\dagger}}\right) =                                                                                               
\log \left(\frac{\rho_{\rm mhd} \kappa_{\rm c,mhd}}{\rho^{\dagger} \kappa_{\rm c}^{\dagger}}\right)
\label{eq:dtaudz}
,\end{eqnarray}

\noindent where again the symbol $^{\dagger}$ is used to indicate that the density
and continuum opacity, which depend on the temperature and gas pressure, have 
been obtained using the different approximations and boundary conditions described in 
Sects.~\ref{sec:hydeq} and ~\ref{sec:mhseq}. As this figure shows, the $z$-spacing between 
discrete $\tau_{\rm c}$ points is better retrieved in the magnetohydrostatic case than 
in the hydrostatic one. We can also see that the differences caused by using different
boundary conditions for $P_{\rm g}(z_{\rm max})$ (red vs. blue lines) are small compared 
to the differences produced by switching from hydrostatic to magnetohydrostatic
equilibrium (solid vs. dashed lines).

\begin{figure*}
    \begin{center}
      \begin{tabular}{cc}
        \includegraphics[width=8cm]{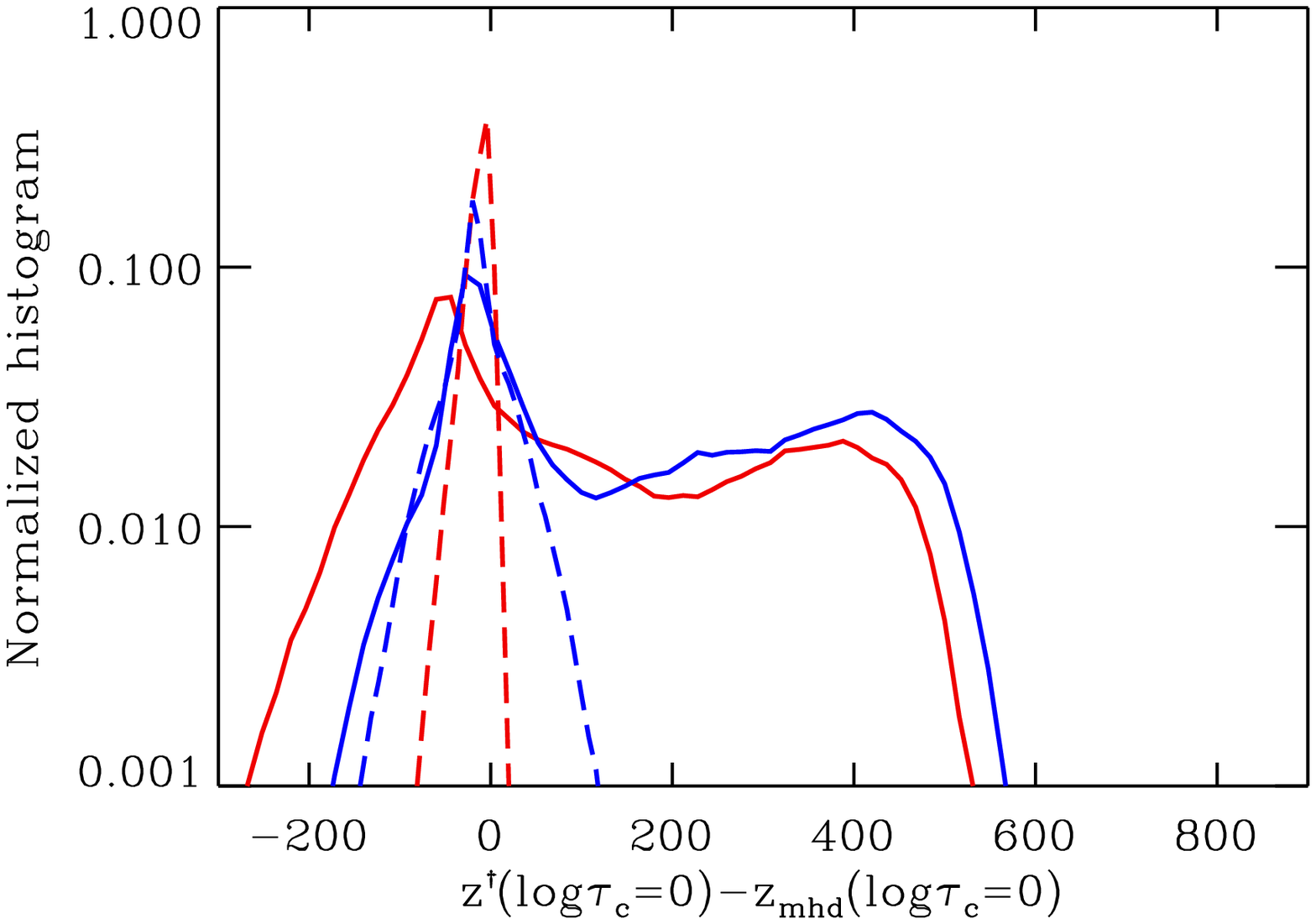} &
        \includegraphics[width=8cm]{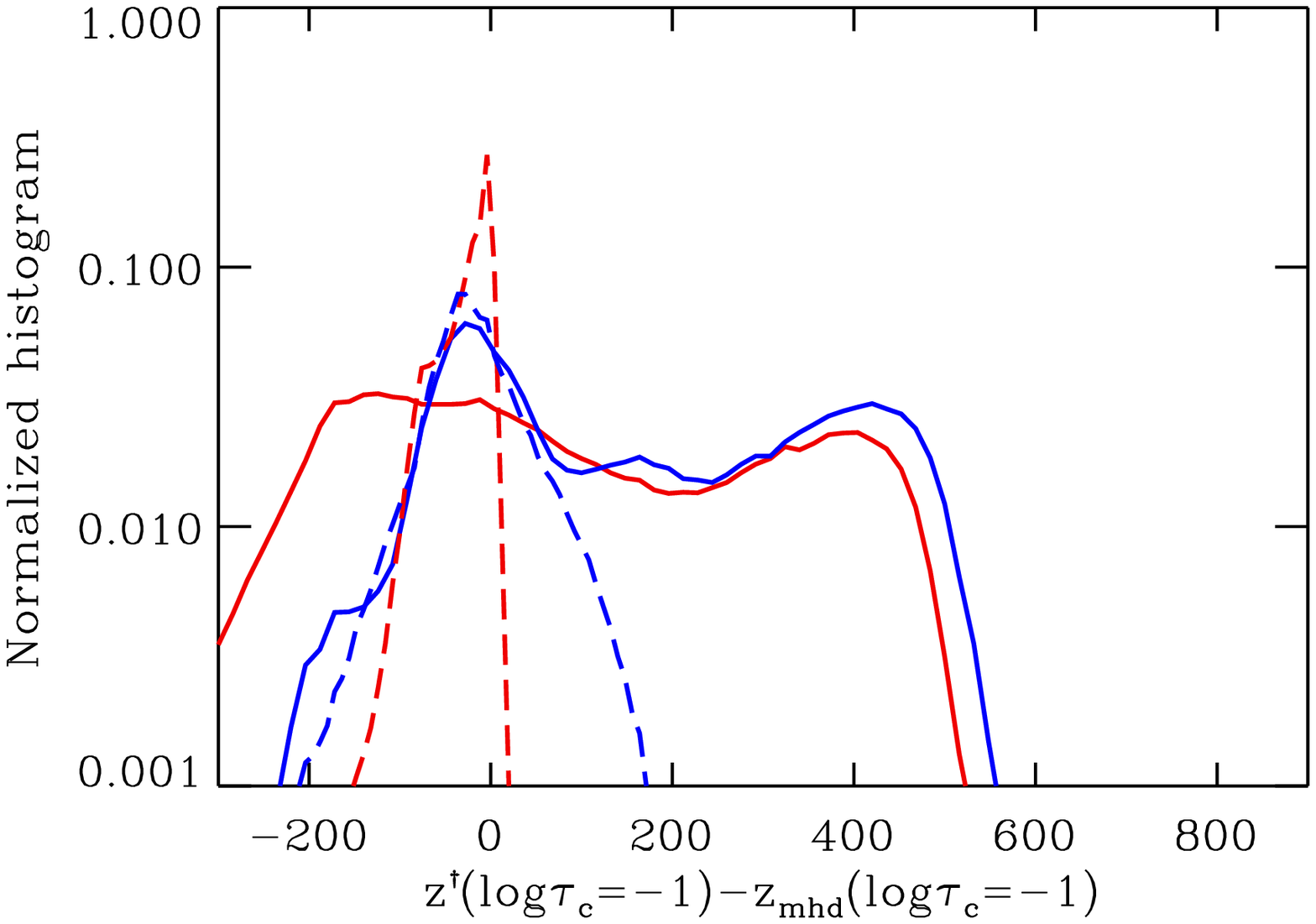} \\
        \includegraphics[width=8cm]{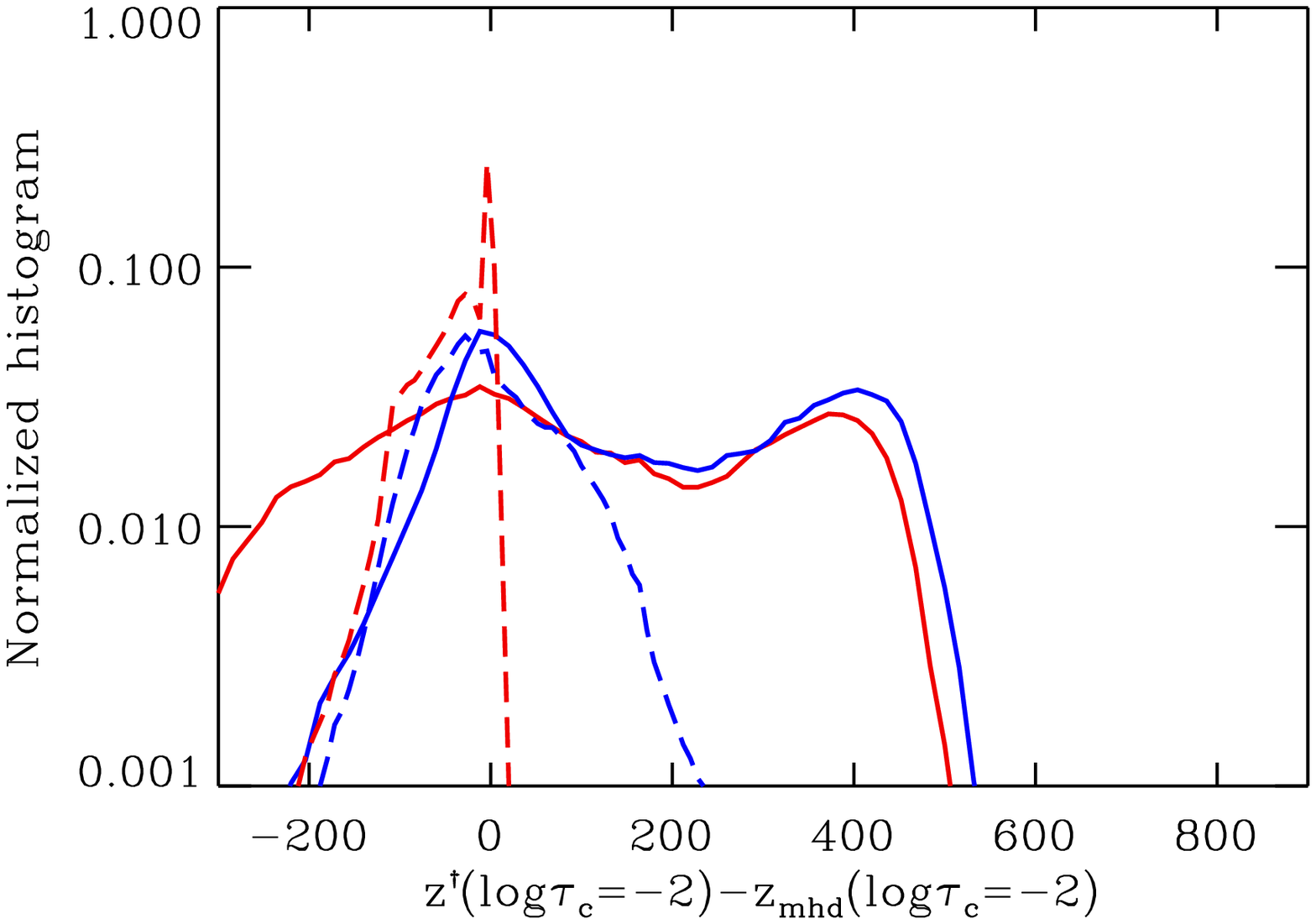} &
        \includegraphics[width=8cm]{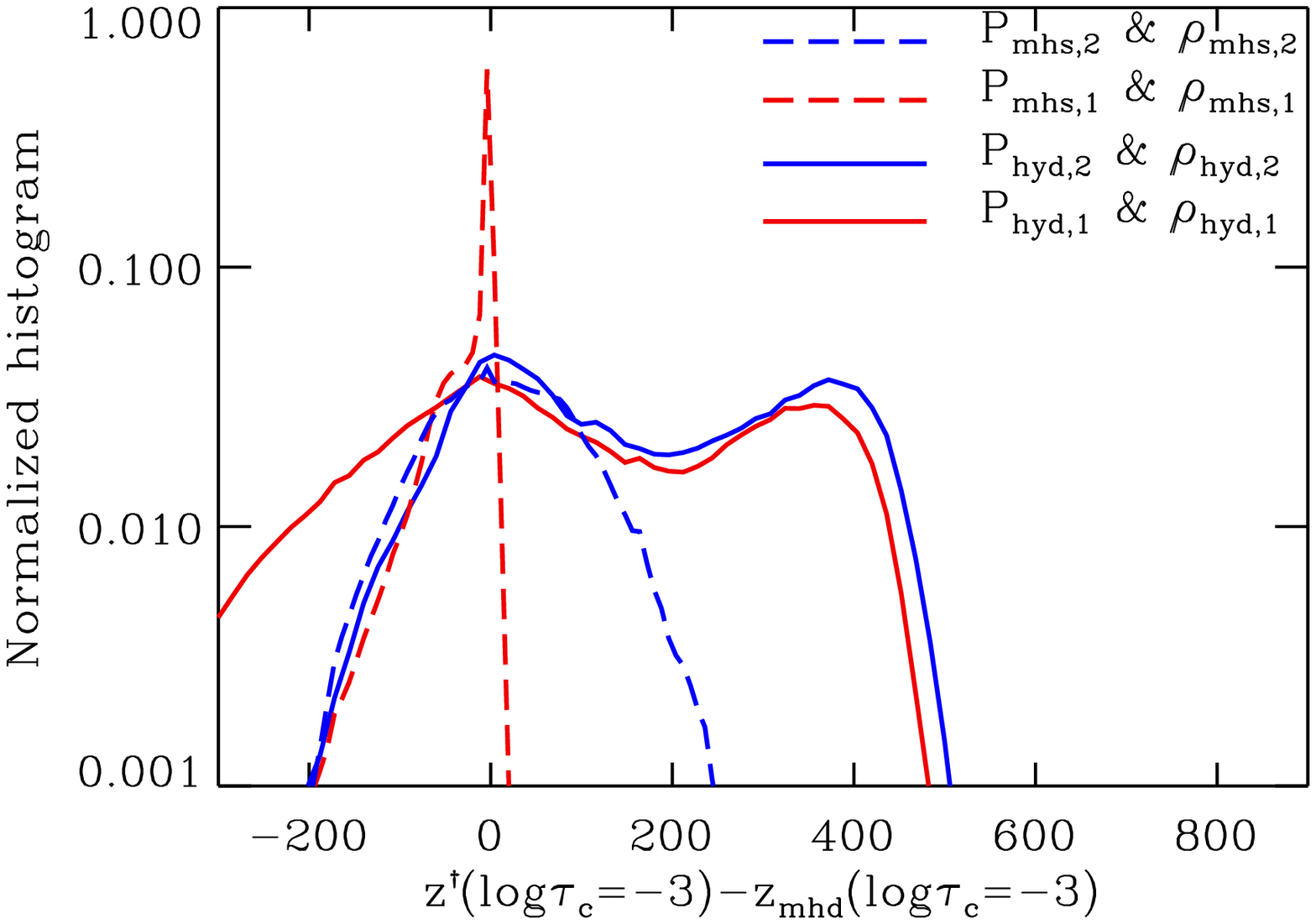}
      \end{tabular}
      \caption{Histograms of the difference between the inferred height $z^{\dagger}$ and the height in the MHD
        simulations $z_{\rm mhd}$ for different optical-depth levels: $\log\tau_{\rm c}=0$ (upper-left), $-1$ (upper-right),
        $-2$ (bottom-left), and $-3$ (bottom-right). Solid colored lines represent the inferences in the
        hydrostatic case (red for boundary condition $P_{\rm hyd,1}$; blue for boundary condition $P_{\rm hyd,2}$),
        and dashed colored lines display the magneto-hydrostatic case (red for boundary condition $P_{\rm mhs,1}$;
        blue for boundary condition $P_{\rm mhs,2}$). See text for details.\label{fig:zhist}}
    \end{center}
\end{figure*}

Once the local stretching or shrinking of the $z-\tau_{\rm c}$ scale has been studied under different 
approximations we can address the full $z-\tau_{\rm c}$ conversion including the effects of the global
shift in this scale produced by the choice of boundary conditions. Figure~\ref{fig:zhist} shows histograms of the
quantity: $\zda(\log\tau_{\rm c})-z_{\rm mhd}(\log\tau_{\rm c})$ at four optical-depth values:
$\log\tau_{\rm c}=[0,-1,-2,-3]$ (upper-left, upper-right, bottom-left, bottom-right). Here, $\zda(\log\tau_{\rm c})$
represents the $z-\tau_{\rm c}$ conversion obtained from the same four tests described above. In other words,
Fig.~\ref{fig:zhist} displays the histograms of the differences between the maps displayed in
Figs.~\ref{fig:zhydeq} and~\ref{fig:zmhseq}, and the same map from the MHD simulations: Fig.~\ref{fig:zmhd}.
The histograms of $\zda-z_{\rm mhd}$ using hydrostatic equilibrium (solid red and solid blue lines) feature
a bimodal distribution at all four optical depths studied. The first of the two peaks of the distribution
is located at $\zda-z_{\rm mhd} \approx 0$ to $-100$ km and is composed by grid points in the granulation
that surrounds the sunspot. The second peak is due mostly to grid points in the sunspot umbra as it is located around
$\zda-z_{\rm mhd} \approx 400$ km. The mean of the absolute value of the differences at each optical-depth
level, $\overline{\|\Delta z\|}$, is in the range $160-200$ km (see Table~\ref{tab:summary}).\\

Histograms of $\zda-z_{\rm mhd}$ using magnetohydrostatic equilibrium (color-dashed lines)
feature a clear single-peak distribution centered around $\zda-z_{\rm mhd} \approx 0$ km,
and yield a value of $\overline{\|\Delta z\|}$ that is a factor three to five better than the hydrostatic case
(see Table~\ref{tab:summary}). The mean of the absolute value of the differences,
$\overline{\|\Delta z\|}$, ranges $10-40$ km in the best-case scenario, whereas it is about $30-70$ km in the practical scenario (see
Table~\ref{tab:summary}). We reiterate here that the \emph{best-case scenario} helps us to understand how far or close the MHD
simulations are to magneto-hydrostatic equilibrium.\\

The width of the histograms around the center value is smaller in the deep photosphere ($\log\tau_{\rm c}=0$;
upper-left panel in Fig.~\ref{fig:zhist}) than in the upper photosphere ($\log\tau_{\rm c}=-3$; lower-bottom panel
in Fig.~\ref{fig:zhist}). This is because the approximation of magneto-hydrostatic equilibrium starts
to break down in the upper photosphere, where the velocity terms that we neglect in Eqs.~\ref{eq:mhseqvec}
and ~\ref{eq:mhspois} (see also description in Sect.~\ref{sec:mhseq}) begin to play a non-negligible role.
Because in this case the histograms feature a single peak that is centered around $\zda-z_{\rm mhd} \approx 0$ km
it is possible to ascribe the width of the histograms to a measure of the standard deviation of the distribution, hence
to an error in the determination of $z-\tau_{\rm c}$ conversion. We refer to this \emph{error} as $\sigma_{z,\tau_{\rm c}}$. Their
values are summarized in Table~\ref{tab:summary} and, as it can be seen, they are comparable to $\overline{\|\Delta z\|}$.
The numbers presented in Tables~\ref{tab:tabpreden} and ~\ref{tab:summary} allow us to establish that in the
magneto-hydrostatic case, the more practical
scenario where the boundary conditions are not known but are instead guessed (i.e interpolated), the uncertainty in the
determination of the $z-\tau_{\rm c}$ conversion increases by a factor of two to three with respect to an ideal scenario where the
boundary conditions are fully known. However, in the hydrostatic case there is very little difference between
both sets of boundary conditions.\\

\begin{table*}
\begin{center}
\caption{Summary of results for the determination of the $z-\tau_{\rm c}$ conversion\label{tab:summary}}
\begin{tabular}{ccccccc|cccc}
  &  & & & $\overline{\|\Delta z\|}$ & & & & $\sigma_{z,\tau_{\rm c}}$ & & \\
\hline
Equation & $z$-BC & $xy$-BC & $\tau_{\rm c}=1$ & $\tau_{\rm c}=10^{-1}$ & $\tau_{\rm c}=10^{-2}$ & $\tau_{\rm c}=10^{-3}$ &
$\tau_{\rm c}=1$ & $\tau_{\rm c}=10^{-1}$ & $\tau_{\rm c}=10^{-2}$ & $\tau_{\rm c}=10^{-3}$ \\
\hline
Hydrostatic & mhd & na & 164.0 & 187.7 & 189.1 & 183.1 & na & na & na & na \\
Hydrostatic & interpol & na & 194.0 & 199.3 & 198.9 & 197.3 & na & na & na & na \\
MHS & mhd & mhd & 11.4 & 27.6 & 41.8 & 21.3 & 12.7 & 30.1 & 42.3 & 37.5 \\
MHS & interpol & interpol & 29.7 & 46.5 & 59.0 & 68.8 & 38.1 & 59.9 & 74.9 & 84.5 \\
\hline
 & & & [km] & [km] & [km] & [km] & [km] & [km] & [km] & [km]\\ 
\end{tabular}
\end{center}
\tablefoot{See Table~\ref{tab:tabpreden} for details.}
\end{table*}

An additional test that we carried out involves the magneto-hydrostatic case (Eq.~\ref{eq:mhspois}) with Neumann boundary conditions for the upper-most
layer of the three-dimensional domain $z_{\rm max}$, while keeping a Dirichlet condition at $z_{\rm min}$. In this case,
the boundary conditions in Eq.~\ref{eq:bcext} were substituted by:

 \begin{equation}
    \left\{
    \begin{tabular}{c}
      $P_{\rm g,mhs}(\xas,y,z) = P_{\rm g,int}(\xas,y,z)$ \\ $P_{\rm g,mhs}(x,\yas,z) = P_{\rm g,int}(x,\yas,z)$ \\
      $P_{\rm g,mhs}(x,y,z_{\rm min}) = P_{\rm g,int}(x,y,z_{\rm min})$ \\
      $\frac{\partial P_{\rm g,mhs}(x,y,z)}{\partial z} \Big\|_{z_{\rm max}} =0.$
    \end{tabular}\right.
    \label{eq:bcext_new}
 \end{equation}

Results in this case are almost identical to those employing Eq.~\ref{eq:bcext}. Density
and pressure stratifications become smoother close to $z_{\rm max}$ compared to those presented
as $\rho_{\rm mhs,2}$ and $P_{\rm mhs,2}$ in Fig.~\ref{fig:mhseq} (blue-dashed lines). However, these
changes are only minor, and in fact Figs.~\ref{fig:zmhseq},~\ref{fig:predenhist}, and ~\ref{fig:zhist},
as well as Tables~\ref{tab:tabpreden} and ~\ref{tab:summary} remain almost unchanged.

\section{Limitations and implementation in Stokes inversions codes}
\label{sec:limitations}

There are some important limitations in the method we have described to obtain reliable
values for the gas pressure, density, and the conversion from $z$ to $\tau_{\rm c}$.
The first limitation has to do with the knowledge of the Lorentz force term in Eq.~\ref{eq:mhseqvec}.
We have assumed that this term can be calculated without any issues because the
magnetic field $\ve{B}$ is fully provided by the numerical simulations (Sect.~\ref{sec:3dsimul}).
However, whenever $\ve{B}$ is inferred from the inversion of the Stokes vector we must
consider that in fact $\ve{B}$ is affected by what is referred to as the
\emph{180 degree ambiguity}. This ambiguity implies that the inversion cannot distinguish
between two possible solutions: $\ve{B}=(B_{\rm x},B_{\rm y},B_{\rm z})$ and $\ve{B}^{\dagger}=(-B_{\rm x},-B_{\rm y},B_{\rm z}),$
at each $(x,y,z)$ grid point in the observed domain. A number of methods
have been developed \citep{metcalf1994,manolis2005,metcalf2006} in the past to address this issue.
If our method is to be applied to actual inferences of the magnetic field via the inversion of the
radiative transfer equation, then we must first solve the 180 degree ambiguity problem.
Failing to do so will certainly return unrealistic values of the electric current, $\ve{j} \propto \ve{\nabla} \times \ve{B}$,
and thus also negatively affect the right-hand side of Eqs.~\ref{eq:mhseqvec} and ~\ref{eq:mhspois}.\\

The second limitation has to do with the fact that even after correctly solving the
180 degree ambiguity problem mentioned above, throughout this paper we consider
that $T(x,y,z)$ and $\ve{B}(x,y,z)$ are known, for instance via the application of our
recently developed Stokes inversion code in the $z$-scale \citep{adur2019invz}. With this
we have shown that it is possible to obtain accurate values for the density $\rho(x,y,z)$
and gas pressure $P_{\rm g}(x,y,z)$ by applying magneto-hydrostatic constraints (Sect.~\ref{sec:mhseq}).
However, one of the main conclusions of \cite{adur2019invz} is that $T$ and $\ve{B}$
can only be properly retrieved by the inversion if $P_{\rm g}$ (or $\rho$) is
reliably known (see also Sect.~\ref{sec:intro}). This problem can only be
solved iteratively, that is, proposing an initial solution for $P_{\rm g}$ and $\rho$
that is used during the Stokes inversion to obtain $T$ and $\ve{B}$. The latter two
physical parameters can then be employed to apply the magneto-hydrostatic constraints
and obtain a better estimation of $P_{\rm g}$ and $\rho$ that is then sent back into the
Stokes inversion. It remains to be proven whether or not this procedure would converge.\\

An additional limitation that is worth mentioning at this point but is not addressed
in this paper concerns the fact that our method works best in the photosphere where the
assumption of magneto-hydrostatic equilibrium is adequate. Higher up, in particular in the chromosphere,
this assumption will surely break down because velocity terms
play an important role in the force balance. The main problem here is that Stokes inversion
codes use the Doppler effect to retrieve the line-of-sight component of the
velocity ($v_{\rm z}$ if at disk center) only and therefore those additional terms cannot
be readily evaluated. A possible solution could be to use time-resolved spectropolarimetric
observations to infer $v_{\rm x}$ and $v_{\rm y}$ \citep{welsch2004vxvy,andres2017vxvy}.\\

Another important limitation that needs to be addressed when applying this method to
real observations is the fact that, unlike numerical simulations, the inversion of
the radiative transfer equation provides the temperature $T$ and magnetic field
vector $\ve{B}$ with different degrees of certainty at each $(x,y,z)$ grid point,
in particular along the $z$-coordinate, with errors increasing quickly below
$\tau_{\rm c}=1$ and above $\tau_{\rm c}=10^{-4}$ (actual values will depend on the
spectral lines employed in the inversion). How the inaccuracies
in the determination of these two physical parameters affect the retrieval of the gas pressure
$P_{\rm g}$ and density $\rho$ remains to be seen. If this effect is too large, extrapolation of the physical quantities
or perhaps a switch to hydrostatic equilibrium (i.e., by making the term related to the magnetic
field on the right-hand side of Eq.~\ref{eq:mhspois} vanish) outside the region where the spectral
lines are sensitive might be needed.

\section{Conclusions}
\label{sec:conclu}

Here we present a new method to determine the gas pressure and density in the solar
atmosphere under the assumption of magneto-hydrostatic
equilibrium. The method was developed to be  used in conjunction with
an inversion code for the polarized radiative transfer equation. Therefore, it considers that the
temperature $T$ and magnetic field $\ve{B}$ are known (i.e., given by the inversion) in the
three-dimensional domain $(x,y,z)$. The proposed method has been tested with a three-dimensional
numerical simulation of a sunspot, and we confirm its potential to retrieve values for the density
and gas pressure that are, in more than 80\% of the grid points in the domain, within
one order of magnitude of the values of the numerical simulation. Moreover, in more than 50\%
of the domain the retrieval is within a factor of two of the numerical simulation
(see Fig.~\ref{fig:predenhist} and Table~\ref{tab:tabpreden}). In contrast, we find that
the approach based on hydrostatic equilibrium (as extensively used in current inversion codes
such as SIR, SPINOR, NICOLE, and SNAPI) determines the correct order of magnitude of these
two physical parameters in only about 45\% of the domain, whereas a more accurate
inference, within a factor of two, occurs only in about 20\% of the domain.\\

Once the pressure and density are known, it is possible to (together with temperatures
from the inversion) calculate a $z-\tau_{\rm c}$ conversion employing Eq.~\ref{eq:ztau}. Again
we have shown that the application of the magneto-hydrostatic solution dramatically
improves this conversion compared to the hydrostatic case. Under ideal conditions,
when taking as boundary conditions the values of the gas pressure provided by the
MHD simulation, our method retrieves the geometrical height $z$ for different $\tau_{\rm c}$-levels
with an accuracy of some $10-40$ km. If instead we employ boundary conditions that
are more adequate for real applications (i.e., boundary conditions obtained from interpolated
models) the error in the determination of the $z$ scale at various $\tau_{\rm c}$-levels
increases to about $30-70$ km (see Table~\ref{tab:summary}).\\

At this juncture it is important to mention that while the method we present here
works better with an inversion code for the radiative transfer equation that retrieves
the physical parameters in $z$ \citep{adur2019invz}, it could also in principle be adapted to
inversion codes that retrieve the physical parameters in $\tau_{\rm c}$. Therefore, the
aforementioned codes that employ hydrostatic equilibrium need not continue to do so.\\

The most obvious advantage of being able to obtain a reliable $z-\tau_{\rm c}$ conversion
is the ability to determine accurate spatial derivatives of the magnetic field and thus
also accurate electric currents \citep{puschmann2010j}, which are considered as proxies
of magnetic reconnection and chromospheric and coronal activity \citep{priest1999}. In
addition, the ability to infer accurate values for the gas pressure and density in
the lower solar atmosphere can significantly help to improve methods that extrapolate the
magnetic field observed in the photosphere towards the chromosphere and corona \citep{wiegelmann2018}.\\

The presented method has however a number of limitations such as a limited applicability
in the upper solar atmosphere (i.e., chromosphere), where the spatial and temporal derivatives
of the velocity play an important role in the momentum equation. It also remains to be
seen how this method performs with real observations, where the spatial resolution is
much lower than in the numerical simulations, and where the determination of the temperature
and magnetic field comes with an attached uncertainty level as a consequence of the
application of the inversion process applied to spectropolarimetric observations. We will
try to address some of these issues in the future.

\begin{acknowledgements}
  This work was supported by the \emph{Deut\-sche For\-schungs\-ge\-mein\-schaft, DFG\/}
  project number 321818926. JMB acknowledges financial support from the Spanish Ministry of Economy
  and Competitiveness (MINECO) under the 2015 Severo Ochoa Program MINECO SEV-2015-0548. The authors
  acknowledge the comments and suggestions for improvement by the referee Dr. Ivan Mili\'{c}. This
  material is based upon work supported by the National Center for Atmospheric Research, which is
  a major facility sponsored by the National Foundation under Cooperative Agreement No. 1852977.
  This research has made use of NASA's Astrophysics Data System.
\end{acknowledgements}

\bibliographystyle{aa}
\bibliography{ms}

\end{document}